\pdfoutput=1
\documentclass[cernpreprint, atlasdraft=false, texlive=2023, UKenglish, texmf, orcidlogo]{atlasdoc}
 
\usepackage{todonotes}
\DeclareBoolOption[true]{LANGEDIT}
\LEcontact{Chip Brock brockr@msu.edu}

\usepackage[subcaption, backend=biber]{atlaspackage}
\usepackage{multicol}
\usepackage{multirow}
\usepackage[backref=false]{atlasbiblatex}
 
\usepackage[BSM, jetetmiss]{atlasphysics}
 
\graphicspath{{logos/}{figures/}}
 
\usepackage{ANA-SUSY-2018-13-PAPER-defs}

\AtlasTitle{Search for long-lived, massive particles in events with displaced vertices and multiple jets in pp collisions at $\sqrt{s} = 13$~TeV with the ATLAS detector}
 
\AtlasAbstract{
A search for long-lived particles decaying into hadrons is presented. The analysis uses 139~fb$^{-1}$ of $pp$ collision data collected at $\sqrt{s} = 13$~TeV by the ATLAS detector at the LHC using events that contain multiple energetic jets and a displaced vertex. The search employs dedicated reconstruction techniques that significantly increase the sensitivity to long-lived particles decaying in the ATLAS inner detector. Background estimates for Standard Model processes and instrumental effects are extracted from data. The observed event yields are compatible with those expected from background processes. The results are used to set limits at 95\% confidence level on model-independent cross sections for processes beyond the Standard Model, and on scenarios with pair-production of supersymmetric particles with long-lived electroweakinos that decay via a small $R$-parity-violating coupling. The pair-production of electroweakinos with masses below 1.5~TeV is excluded for mean proper lifetimes in the range from 0.03~ns to 1~ns. When produced in the decay of $m(\tilde{g})=2.4$~TeV gluinos, electroweakinos with $m(\tilde\chi^0_1)=1.5$~TeV are excluded with lifetimes in the range of 0.02~ns to 4~ns.
}
 
\author{The ATLAS Collaboration}
 
\AtlasRefCode{SUSY-2018-13}

\AtlasJournal{JHEP}
\AtlasJournalRef{JHEP 06 (2023) 200}

\PreprintIdNumber{CERN-EP-2023-002}

\hypersetup{pdftitle={ATLAS document},pdfauthor={The ATLAS Collaboration}}
 
\begin{document}
 
\maketitle
 
\section{Introduction}
 
The interest in searches for long-lived particles (LLPs) has recently grown and they have become an important part of the hunt for new physics at the Large Hadron Collider~(LHC). LLPs appear in a variety of models trying to address fundamental puzzles of the Standard Model (SM) of particle physics and yield a large palette of signatures at colliders that inspires a diverse and ambitious search programme at the LHC~\mbox{\cite{Lee_2019,Alimena_2020}.}
 
There are various mechanisms by which particles can obtain significant lifetimes ($\gtrapprox 10$~ps) in beyond the Standard Model (BSM) theories. The decays of such particles can be suppressed in so-called Hidden Valley models where large barrier potentials reduce the rate of kinematically allowed decays~\cite{Strassler:2006im}. Decays happening via a highly virtual intermediate state, or in highly compressed mass spectra can also result in long lifetimes~\cite{ArkaniHamed:2004fb,Giudice:2004tc}. Finally, long-lived particles also appear in models with small couplings, such as those often found in $R$-parity-violating supersymmetry (SUSY)~\cite{Barbier:2004ez}, as happens for the simplified models used as a benchmark for the search presented here.
 
SUSY~\cite{Golfand:1971iw,Volkov:1973ix,Wess:1974tw,Wess:1974jb,Ferrara:1974pu,Salam:1974ig} models have the potential to achieve gauge-coupling unification, provide an explanation for dark matter, and alleviate the naturalness problem~\cite{Sakai:1981gr,Dimopoulos:1981yj,Ibanez:1981yh,Dimopoulos:1981zb}. In the minimal supersymmetric extension to the SM (MSSM)~\cite{Fayet:1976et,Fayet:1977yc}, $R$-parity-violating (RPV) couplings that violate baryon-number and lepton-number conservation naturally occur at tree level. The RPV terms of the MSSM superpotential are given by
 
\begin{equation*}{\mathcal W}_{\textrm{RPV}}=\mu_i \ell_i h_u + \lam_{ijk} \ell_i \ell_j\bar e_k +\lamp_{ijk} \ell_iq_j\bar d_k +\lampp_{ijk}\bar u_i\bar d_j\bar d_k,
\end{equation*}
 
\noindent where $\mu_i$, $\lam_{ijk}$, $\lamp_{ijk}$, and $\lampp_{ijk}$ are the RPV couplings, $\ell$ and $e$ represent the lepton and charged-lepton supermultiplets, $h_u$ represents the up-type Higgs supermultiplets, and $q$, $u$, and $d$ represent the quark, up-type quark, and down-type quark supermultiplets~\cite{Barbier:2004ez}.  The symbols $i$, $j$, and $k$ are generation indices.
 
\begin{figure}[h]
\centering
\begin{subfigure}[b]{0.31\textwidth}
\includegraphics[width=\textwidth]{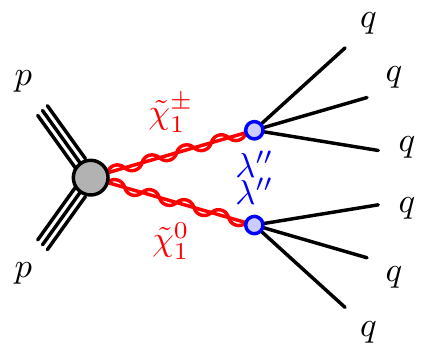}
\caption{}
\label{fig:diagramsb}
\end{subfigure}
\begin{subfigure}[b]{0.31\textwidth}
\includegraphics[width=\textwidth]{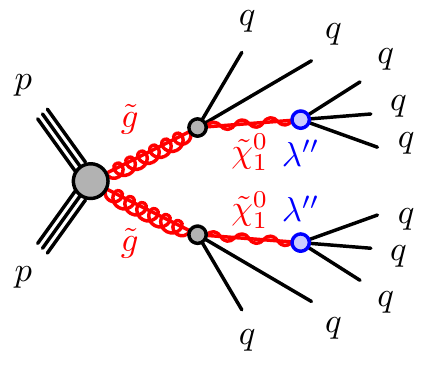}
\caption{}
\label{fig:diagramsa}
\end{subfigure}
\caption{Diagrams showing production of a (a) chargino-neutralino pair (\chinoonepm\ninoone) and (b) a gluino pair (\gluino\gluino), in which each gluino decays into a pair of quarks and a long-lived neutralino (\ninoone). The \chinoonepm and \ninoone decay into three quarks via the R-parity-violating coupling $\lambda^{''}$. For sufficiently small values of $\lambda^{''}$, the lifetime of the \chinoonepm and \ninoone becomes long enough to cause decays which are significantly displaced from their production points.}
\end{figure}

The search presented in this paper targets small values of the \lampp\ couplings, assuming all other RPV couplings to be exactly zero. Several low- and intermediate-energy experimental observations constrain the value of the RPV couplings to be small~\cite{Goity_1995, ZWIRNER1983103}. This naturally leads to suppression of the decay processes and can give rise to long-lived SUSY particles. A nonzero $\lampp$ coupling allows the electroweakinos, supersymmetric states that arise from the mixing of the superpartners of the $U(1)_{Y}$, $SU(2)_{L}$ and Higgs fields, to decay into three quarks. Two different signal scenarios are used as benchmarks in this analysis. The Electroweak (EWK) RPV scenario, shown in \Fig{\ref{fig:diagramsb}}, considers the associated production of the lightest charged and neutral electroweakinos, that is, $\chinoonepm\ninoone$, $\chinoonep\chinoonem$, $\ninotwo\ninoone$ and $\chinoonepm\ninotwo$, assuming pure-higgsino states. In the Strong RPV scenario, shown in \Fig{\ref{fig:diagramsa}}, pairs of gluinos $\gluino$, the superpartners of the SM gluons, are produced. Each \gluino\ promptly decays with a 100\% branching ratio to a SM quark--antiquark pair and a neutralino $\ninoone$, the lightest neutral electroweakino. The electroweakinos in both models decay with a 100\% branching ratio via the $\lampp$ couplings to light-flavour quarks.
 
This paper presents the first search for the RPV models described above in events with a massive, multi-track displaced vertex and multiple energetic jets using proton--proton collisions collected by the ATLAS experiment at a centre-of-mass energy of $\sqrt{s} = 13~\TeV$. Dedicated track and vertex reconstruction algorithms are employed to maximize the reconstruction and selection efficiency of displaced vertices originating from LLP decays which occur inside of the tracking volume but at a discernible distance from the interaction point of the incoming beams. The analysis does not require that the jets are linked to the displaced vertex in order to retain sensitivity to various BSM decay topologies. Signal region (SR) criteria are designed to select on the order of one background event in the available data sample, while maximizing the expected signal yield. The predictions of the background yields from several sources are entirely derived from data, with individual contributions estimated from dedicated control regions (CRs).
 
Similar searches were performed previously by the ATLAS, CMS, and LHCb collaborations at $\sqrt{s} = 7, 8$ and 13~\TeV~\cite{SUSY-2011-02, EXOT-2011-12,SUSY-2014-02,SUSY-2018-33,CMS-EXO-19-013,CMS-EXO-17-018,Aaij_2017}. In this analysis, novel techniques to estimate background processes that can produce a displaced vertex signal were developed, and sensitivity to RPV SUSY models predicting production of long-lived electroweakinos with the ATLAS experiment is demonstrated for the first time.
 
\section{ATLAS detector}
\newcommand{\AtlasCoordFootnote}{
ATLAS uses a right-handed coordinate system with its origin at the nominal interaction point (IP) in the centre of the detector and the \(z\)-axis along the beam pipe. The \(x\)-axis points from the IP to the centre of the LHC ring, and the \(y\)-axis points upwards. Cylindrical coordinates \((r,\phi)\) are used in the transverse plane, \(\phi\) being the azimuthal angle around the \(z\)-axis. The pseudorapidity is defined in terms of the polar angle \(\theta\) as \(\eta = -\ln \tan(\theta/2)\).}
 
The ATLAS detector~\cite{PERF-2007-01,PIX-2018-001} at the LHC is a multipurpose particle detector with a forward--backward symmetric cylindrical geometry and a near \(4\pi\) coverage in solid angle.\footnote{\AtlasCoordFootnote}
It consists of an inner tracking detector (ID) surrounded by a thin superconducting solenoid providing a \SI{2}{\tesla} axial magnetic field, electromagnetic (EM) and hadron calorimeters, and a muon spectrometer (MS).
The inner tracking detector covers the pseudorapidity range \(|\eta| < 2.5\).
It consists of silicon pixel, silicon microstrip (SCT), and transition radiation tracking (TRT) detectors. The pixel detector has four barrel layers and three disks in each of the forward and backward regions. The barrel layers surround the beam pipe at radii of 33.3~mm, 50.5~mm, 88.5~mm, and 122.5~mm, covering \(|\eta| < 1.9\). The pixel detector spans the radius range from 3~cm to 12~cm, the SCT spans 30~cm to 52~cm, and the TRT spans 56~cm to 108~cm.
 
Lead/liquid-argon (LAr) sampling calorimeters provide EM energy measurements with high granularity.
A steel/scintillator-tile hadron calorimeter covers the central pseudorapidity range \(|\eta| < 1.7\).
The endcap and forward regions are instrumented with LAr calorimeters for both the EM and hadronic energy measurements up to \(|\eta| = 4.9\). The MS surrounds the calorimeters and is based on three large air-core toroidal superconducting magnets with eight coils each. The field integral of the toroids ranges between \SI{2}{\tesla\metre} and \SI{6}{\tesla\metre} across most of the detector.
The MS includes a system of precision tracking chambers and fast detector chambers for triggering. A two-level trigger system is used to select events~\cite{TRIG-2016-01}. The first-level trigger is implemented in hardware and uses a subset of the detector information to reduce the accepted rate to at most \SI{100}{\kHz}. This is followed by a software-based trigger that reduces the accepted event rate to \SI{1}{\kHz} on average depending on the data-taking conditions.
An extensive software suite~\cite{ATL-SOFT-PUB-2021-001} is used for real and simulated data reconstruction and analysis, for operation, and in the trigger and data acquisition systems of the experiment.

\FloatBarrier
\section{Data and simulated event samples}
\label{sec:datasets}
 
Proton–-proton collision data recorded by the ATLAS detector between 2015 and 2018 are used for this analysis. In this period, the LHC delivered colliding beams with an average number of $pp$ interactions per bunch crossing of 33.7. After applying beam, detector, and data-quality criteria~\cite{DAPR-2018-01} the total integrated luminosity of the data is 139~\ifb. The uncertainty in the combined 2015–2018 integrated luminosity is 1.7\%, obtained using the LUCID-2 detector for the primary luminosity measurements~\cite{ATLAS-CONF-2019-021,Avoni:2018iuv}.
 
The data were recorded using a combination of triggers requiring multiple energetic jets with transverse momentum \pT\ thresholds that depend on the jet multiplicity. These ranged from four jets with $\pT>100$~\GeV\ to seven jets with $\pT>45$~\GeV.
 
Monte Carlo (MC) simulated event samples are used to aid the estimate of background from SM processes and to model the targeted signal models. The matrix element (ME) calculation for the SUSY production processes was performed to leading-order precision with \MGNLO[2.6.2]~\cite{Alwall:2014hca} interfaced to \PYTHIA[8.212]~\cite{Sjostrand:2014zea} for  parton showering (PS) and hadronization. All other SUSY-particle contributions are assumed to be decoupled. The ME calculation was performed at tree level and included the emission of up to two additional partons for all signal samples. The parton distribution function (PDF) set used for the generation of the signal samples was \NNPDF[2.3lo]~\cite{Ball:2012cx} with the A14~\cite{ATL-PHYS-PUB-2014-021} set of tuned underlying-event and shower parameters. The ME–PS matching was performed with the CKKW-L prescription~\cite{Lonnblad:2001iq,Lonnblad:2011xx}, with a matching scale set to one quarter of the produced SUSY particle mass. The electroweakino production cross sections were computed at next-to-leading-order (NLO) plus next-to-leading-log (NLL) precision~\cite{Beenakker:1999xh,Debove:2010kf,Fuks:2012qx,Fuks:2013vua,Fiaschi:2018hgm}
in a limit of mass-degenerate higgsinos $\ninotwo$, $\chinoonepm$,  and $\ninoone$. The nominal cross-section and the uncertainty were taken from an envelope of cross-section predictions using different PDF sets and factorization and renormalization scales, as described in Ref.~\cite{Borschensky:2014cia}. Signal cross-sections for the gluino pair production were calculated to approximate Next-to-Next-to-Leading-Order  (\NNLO) in the strong coupling constant, adding the resummation of soft gluon emission at Next-to-Next-to-Leading-Log (NNLL) accuracy (approximate \NNLO+NNLL)~\cite{Beenakker:2016lwe,Beenakker:2014sma,Beenakker:2013mva,Beenakker:2011sf,Beenakker:2009ha,Kulesza:2009kq,Kulesza:2008jb,Beenakker:1996ch}. The nominal cross-section and its uncertainty were derived using the PDF4LHC15\_mc PDF set, following the recommendations of Ref.~\cite{Butterworth:2015oua}.  In the electroweakino case, the cross-sections ranged from $300 \pm 10$~fb at $\ninoone$ mass of 300~\GeV\ to $21.5 \pm 3.7$~ab at 1700~\GeV, assuming pure-higgsino states. In the gluino case, they ranged from $8.9 \pm 1.4$~fb at a \gluino\ mass of 1.6~\TeV\  to $46 \pm 14$~ab at 2.6~\TeV.
 
Samples of multijet events were generated using \PYTHIA[8.230] with leading-order matrix elements for dijet production which were matched to the parton shower.
The \NNPDF[2.3lo] PDF set was used in the ME generation, the parton shower, and the simulation of the multi-parton interactions. The A14
set of tuned parameters was used. These samples are used for an independent closure test of the data-driven background estimates.
 
To simulate the effects of additional $pp$ collisions in the same and nearby bunch crossings (pile-up), additional interactions were generated using the soft QCD processes provided by \PYTHIA[8.186] with the A3 tune~\cite{ATL-PHYS-PUB-2016-017} and the MSTW2008LO PDF set~\cite{Martin:2009iq}, and overlaid onto each simulated hard-scatter event. The MC samples were reweighted so that the pile-up distribution matches the one observed in the data.
The MC samples were processed through an ATLAS detector simulation~\cite{SOFT-2010-01} based on \GEANT~\cite{Agostinelli:2002hh}. All MC samples were reconstructed in the same manner as the data.
 
\FloatBarrier
\section{Event reconstruction}
 
The signal processes targeted by this search include massive particles with mean proper lifetimes $\tau$ up to $O(10)$~ns. Several charged decay products will originate from locations in the detector with a substantial displacement from the primary proton--proton interaction and require dedicated techniques to be efficiently reconstructed.
 
The standard ATLAS track reconstruction algorithm is optimized for charged particles originating from the interaction region, or from decays of short-lived particles such as $b$-hadrons. Constraints are set on the transverse and longitudinal impact parameters of track candidates relative to the interaction region ($|d_0|<10$~mm and $|z_0|<250$~mm, respectively), and their hit multiplicities in order to reduce the computational complexity. These selections result in inefficiencies in the reconstruction of tracks with large impact parameters, corresponding to charged particles produced in the LLP decays. To reconstruct these tracks, a dedicated track reconstruction, called \emph{large-radius tracking} (LRT)~\cite{ATL-PHYS-PUB-2017-014}, is performed. This reconstruction step takes as input the hits left over from the standard tracking procedure and applies looser requirements, in particular on the transverse and longitudinal impact parameters ($|d_0| < 300$~mm and $|z_0| < 1500$~mm).
Tracks from the standard processing and the LRT processing are treated collectively in all subsequent steps of reconstruction.
 
Collision vertices (CV) are reconstructed~\cite{ATL-PHYS-PUB-2015-026} with at least two associated tracks that are consistent with originating from the beam collision region in the $x$--$y$ plane. For each reconstructed vertex the quantity $\Sigma_{\pT}$, defined as the sum of squared transverse momenta of associated tracks, is computed. The primary vertex (PV) in the event is the vertex with the highest $\Sigma_{\pT}$.
 
A dedicated secondary-vertex reconstruction algorithm for LLP decays is employed to reconstruct displaced vertices (DVs)~\cite{ATL-PHYS-PUB-2019-013}. As input to this vertex reconstruction, only tracks with transverse momentum $\pT > 1$~\GeV\ and which satisfy several additional requirements on the number and distribution of the associated hits in the tracking detectors are considered. The vertexing starts by forming two-track seed vertices, where at least one track in each two-track seed vertex must have $|d_0| > 2$~mm.
Tracks used to seed vertex formation are called \textit{selected tracks}. The two-track seed vertices are then iteratively merged to form $n$-track vertices. Ambiguities due to tracks being compatible with more than one vertex are resolved by comparing their $\chi^2$ values in the vertex fits, or by merging the vertices if the distance between their estimated positions is not significant enough. As a final step, additional tracks are attached to the reconstructed vertices in order to recover inefficiencies from the vertex seeding selection. These tracks are referred to as \textit{attached tracks} and have looser hit selection requirements than selected tracks.
The resulting DV reconstruction efficiency, defined as the probability for a true neutralino decay to be matched with a reconstructed DV as a function of the radial position R, and the total vertex selection efficiency are shown in Figure~\ref{fig:DVefficiency}, highlighting the improvements due to LRT and attached tracks. The efficiency degrades rapidly beyond 300~mm due to a requirement on the minimum number of hits ($N_\text{hits}\geq7$) applied in the LRT algorithm. The drop in efficiency at small radii ($R\lessapprox5$~mm) is a result of the $|d_0|$ requirement on the two-track seed vertices. The drops in efficiency at larger radii in Figure~\ref{fig:DVefficiencyb} occur due to the baseline DV selection, defined in Section~\ref{sec:selection}, which vetoes DVs reconstructed in regions of detector material. 
Larger neutralino masses lead to more high-\pT\ particles being produced in the decay, which increase the reconstruction efficiency of the DV.

\begin{figure}[t!]
\centering
\begin{subfigure}[b]{0.45\textwidth}
\includegraphics[width=\textwidth]{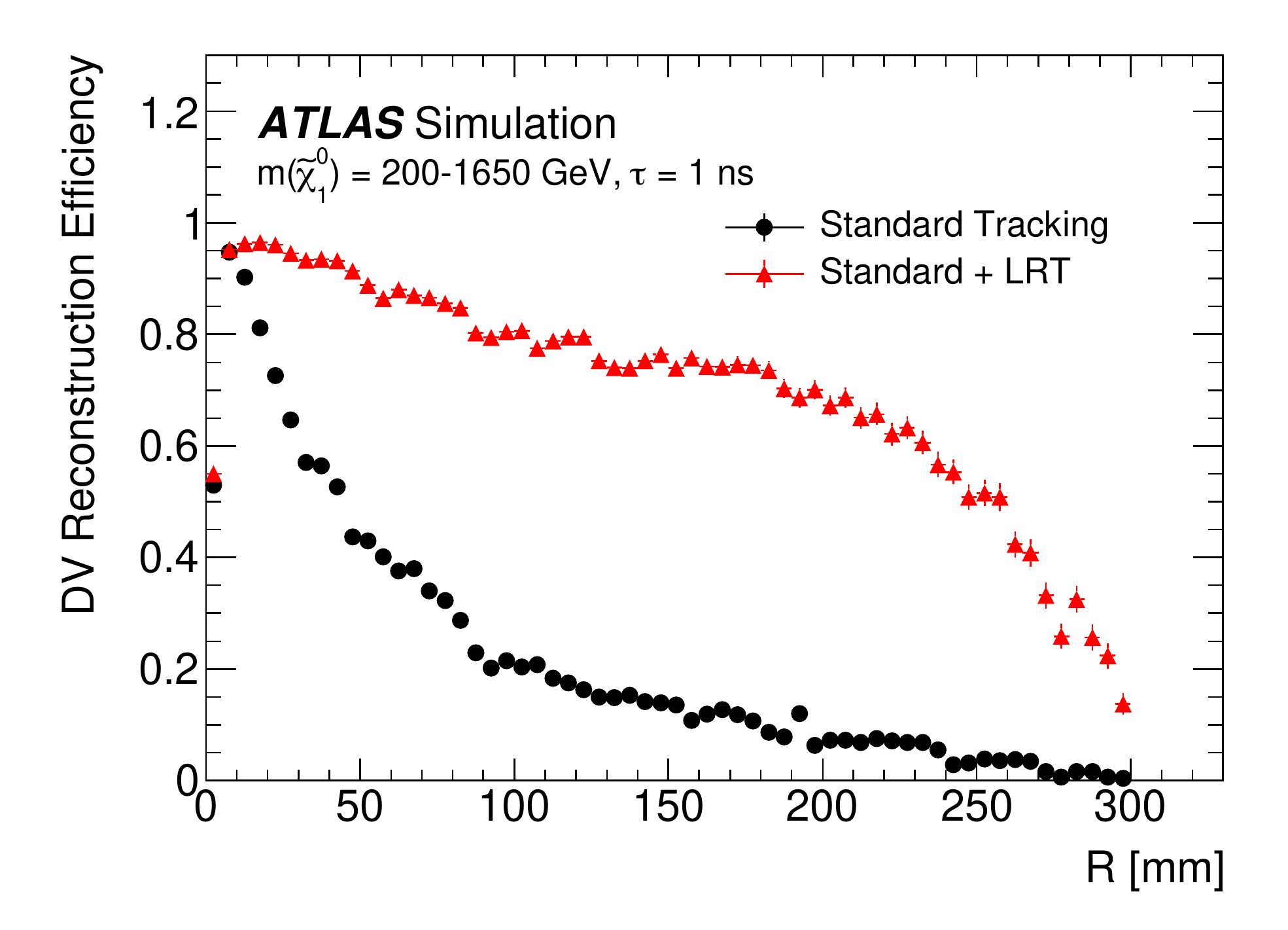}
\caption{}
\end{subfigure}
\begin{subfigure}[b]{0.45\textwidth}
\includegraphics[width=\textwidth]{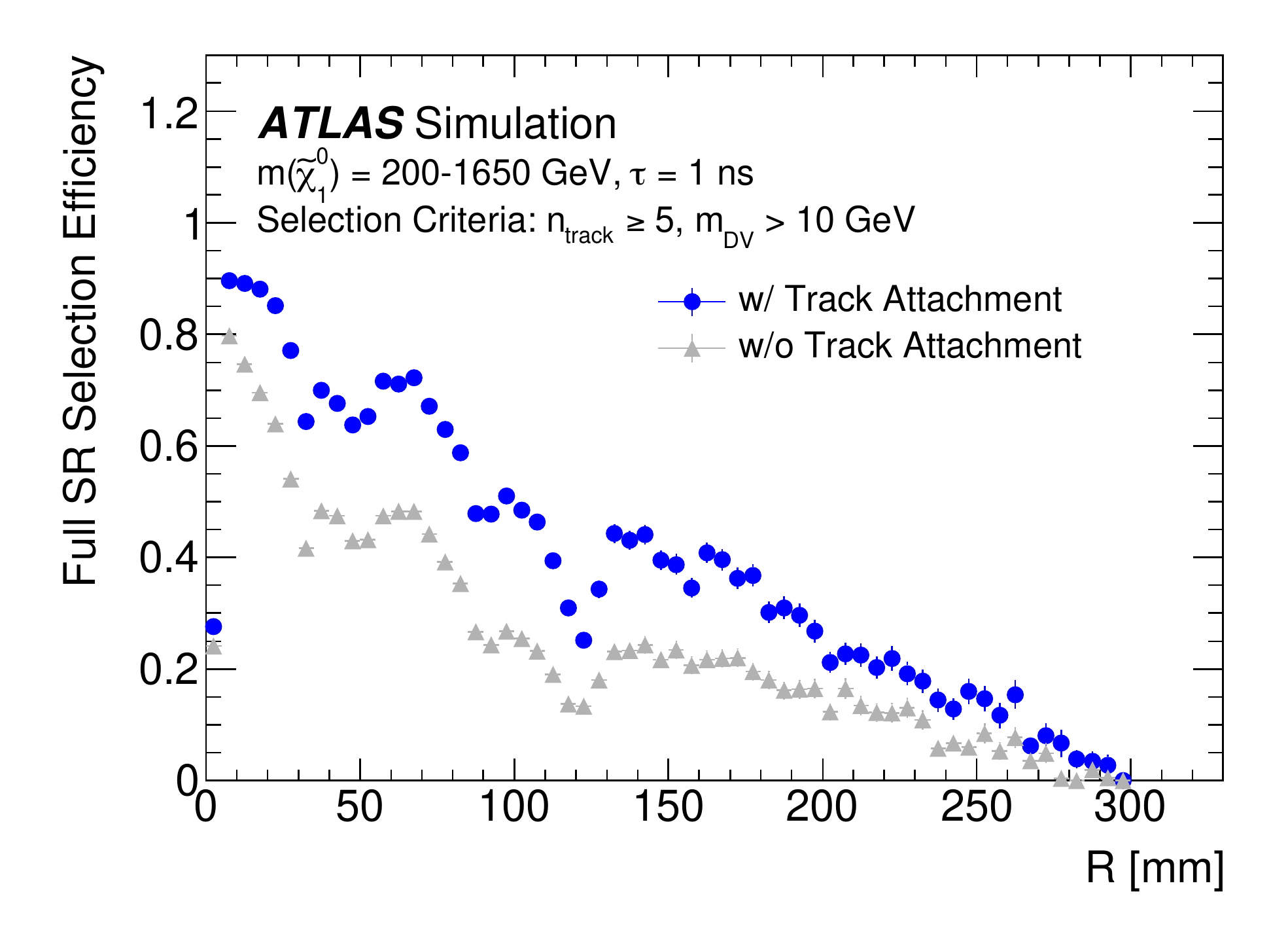}
\caption{}
\label{fig:DVefficiencyb}
\end{subfigure}
\caption{In (a), the DV reconstruction efficiencies with and without the special LRT processing are shown for a combined sample of simulated Strong RPV and EWK RPV events with $m(\tilde{g})=1.8 - 2.2$~\TeV\ , $m(\ninoone)=200-1650$~\GeV\, and $\tau(\ninoone)=1.0$~ns. The efficiency is defined as the probability for a true neutralino decay with at least two charged decay products with $\pt>1$ \GeV\ and $|\eta|<5$ to be matched with a reconstructed DV. In (b), the impact of attached tracks on the total SR selection efficiency of the same combined sample is shown. The total SR selection efficiency is defined in Section~\ref{sec:selection}. }
\label{fig:DVefficiency}
\end{figure}
 
While electrons and muons are not used in the event selection in this search, they are used in a procedure to remove overlapping objects and define the final collection of jets. Muon candidates are reconstructed in the region $|\eta|<2.7$ from MS tracks matching ID tracks.
Candidate muons are required to have $\pT>\SI{10}{\GeV}$ and satisfy the \textit{medium} identification requirements defined in Ref.~\cite{MUON-2018-03}, based on the number of hits in the different ID and MS subsystems, and on the ratio  of the charge and momentum ($q/p$) measured in the ID and MS divided by the sum in quadrature of their corresponding uncertainties.
 
Electron candidates are reconstructed from isolated electromagnetic calorimeter energy deposits matched to ID tracks and are required to have $|\eta|<2.47$, a transverse momentum $\pT>\SI{10}{\GeV}$, and to satisfy the \textit{LooseAndBLayer} requirement defined in Ref.~\cite{PERF-2017-01}, which is based on a likelihood using measurements of shower shapes in the calorimeter and track properties in the ID as input variables. Electrons which share an ID track with a muon are discarded.
 
Jets are reconstructed from three-dimensional energy clusters in the calorimeters~\cite{ATL-PHYS-PUB-2015-036} using the anti-$k_t$ jet clustering algorithm~\cite{Cacciari:2008gp, Fastjet} with a radius parameter $R=0.4$. Only jet candidates with $\pT>\SI{20}{\GeV}$ and $|\eta|<2.8$ are considered. Jets are calibrated using MC simulation with corrections obtained from in situ techniques~\cite{PERF-2016-04}.
Events are discarded if they contain any jet with $\pT>\SI{20}{\GeV}$ not satisfying basic quality selection criteria designed to reject detector noise and non-collision backgrounds by imposing the \textit{BadLoose} cleaning selection, described in Ref.~\cite{ATLAS-CONF-2015-029}, without the cuts on the fraction of jet energy deposited within the electromagnetic calorimeter and the jet charged-fraction. Jets within an angular distance $\Delta R =\sqrt{(\Delta \eta)^2+(\Delta\phi)^2} = 0.2$ of an electron candidate are discarded.
 
A second collection of jets is reconstructed to aid in estimating the SM background by exploiting the track density in an event. These jets, referred to as track-jets, are constructed with an anti-$k_t$ algorithm with $R = 0.4$ using all tracks with $\pT > 1$ \GeV\ and $|d_0|<2$~mm. In contrast to the jet reconstruction algorithm from calorimeter energy deposits, the track jet reconstruction is not sensitive to displaced LLP decays, due to the selection on the tracks $|d_0|$. In order to take into account the effects arising from the presence of pile-up interactions, track-jets are reconstructed for each collision vertex in the event. Before track jet-finding, each track is matched with only one reconstructed collision vertex to minimize the probability of clustering together tracks which originate from different interactions into a single jet. The matching proceeds as follows: first, tracks are assigned with a collision vertex if their longitudinal impact parameter significance satisfies $|z_0/\sigma_{z_0}| < 3$ relative to that vertex. Second, any remaining unassociated tracks are assigned to a collision vertex if their longitudinal impact parameter satisfies $|z_0| < 0.5$~mm relative to that vertex. If a track is found to be compatible with multiple collision vertices, then the vertex with the highest $\Sigma_{\pT}$ is chosen. Track-jet finding is then done for each collision vertex. Track-jets are retained for further analysis if their \pT\ satisfies $\pT > 20$ \GeV.
 
\FloatBarrier
\section{Event selection}
\label{sec:selection}
 
The LRT reconstruction algorithm used in this search is computationally expensive and so it was not run on all events recorded with the ATLAS detector. Instead, a selection referred to as a \textit{filter} is applied to the events satisfying the trigger conditions described in Section~\ref{sec:datasets}. The events passing the filter requirements compose a special data stream that is processed with the LRT algorithm. This analysis uses two filters which apply a selection on uncalibrated jets: the \textit{Trackless jet} and the \textit{High-\pT\ jet} filters. The Trackless jet filter targets the electroweakino pair production signals and the High-\pT\ jet filter targets the gluino pair production signals.
The Trackless jet filter selects events with a high multiplicity of jets ($>3$) satisfying \pT thresholds ranging from 45 to 100~\GeV. Furthermore, events are required to have at least one Trackless jet with $\pT > 70$~\GeV\ and $|\eta|<2.5$, or at least two Trackless jets with $\pT > 50$~\GeV\ and $|\eta|<2.5$. A jet is considered trackless if the sum of the transverse momenta of all standard (non-LRT) tracks associated with the jet is less than 5~\GeV.
The High-\pT\ jet filter selects events with a high multiplicity of jets ($>3$) satisfying \pT thresholds ranging from 75 to 220 \GeV. There is no requirement on the presence of Trackless jets in the High-\pT\ jet filter.
 
An offline jet \pT selection is applied after jet calibration. The specific choices are made  for the \pT\ thresholds to ensure a full efficiency of the offline event selection relative to the trigger and filter requirements. These thresholds depend on the jet multiplicity and range from 55 to 137~\GeV\ for the events passing the Trackless jet filter and from 90 to 250~\GeV\ for the events passing the High-\pT\ jet filter.
 
Candidate events are required to have at least one reconstructed collision vertex and satisfy the High-\pT\ jet or Trackless jet selection requirements.
 
DVs passing the baseline selection must be reconstructed with a high goodness-of-fit, $\chi^2/n_\textrm{DoF} < 5$, to reject vertices resulting from misreconstruction. The reconstructed vertex must be within a fiducial region, which has an outermost boundary defined as $R_\text{DV}=\sqrt{x_\text{DV}^2+y_\text{DV}^2}<300$~mm and $|z_\text{DV}|<300$~mm, where $x_\text{DV}$, $y_\text{DV}$, and $z_\text{DV}$ are the cartesian coordinates of the DV. In addition, the DV must be at least $4$~mm away from any collision vertex in the transverse plane, both to avoid confusing the DV with one of those vertices and also to reduce the heavy-flavour background which could otherwise enter the signal region. To reduce background from hadronic interactions, DVs must satisfy a \textit{material map} veto, which rejects DVs which have a reconstructed position consistent with the position of detector material. The material map, shown in Figure~\ref{fig:matmap}, is constructed using the known positions of detector elements and from the positions of low-mass vertices reconstructed in data. The material map veto removes 48\% of the fiducial volume.
 
\begin{figure}[htb]
\centering
\begin{subfigure}[b]{0.45\textwidth}
\includegraphics[width=\textwidth]{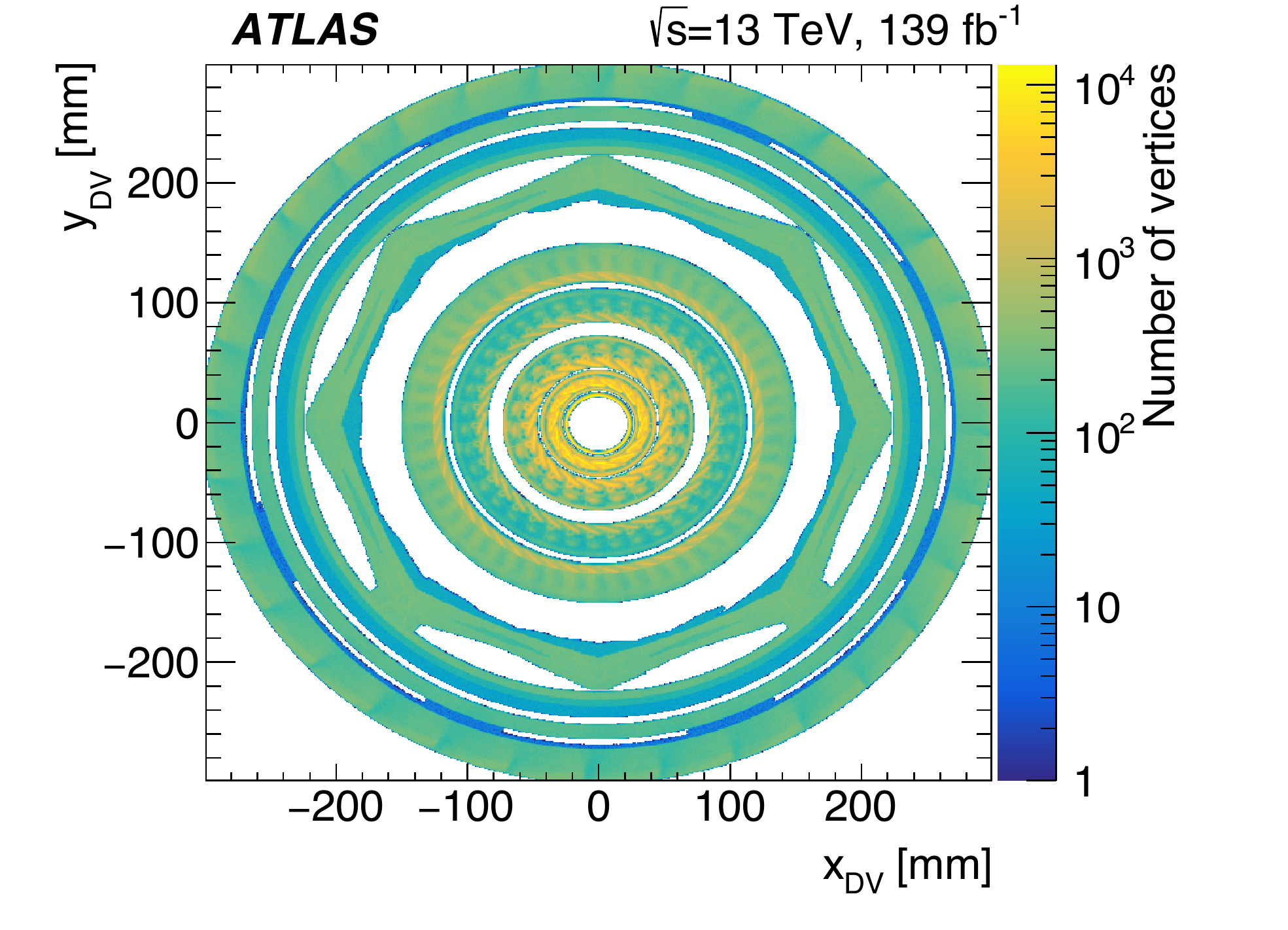}
\caption{}
\end{subfigure}
\begin{subfigure}[b]{0.45\textwidth}
\includegraphics[width=\textwidth]{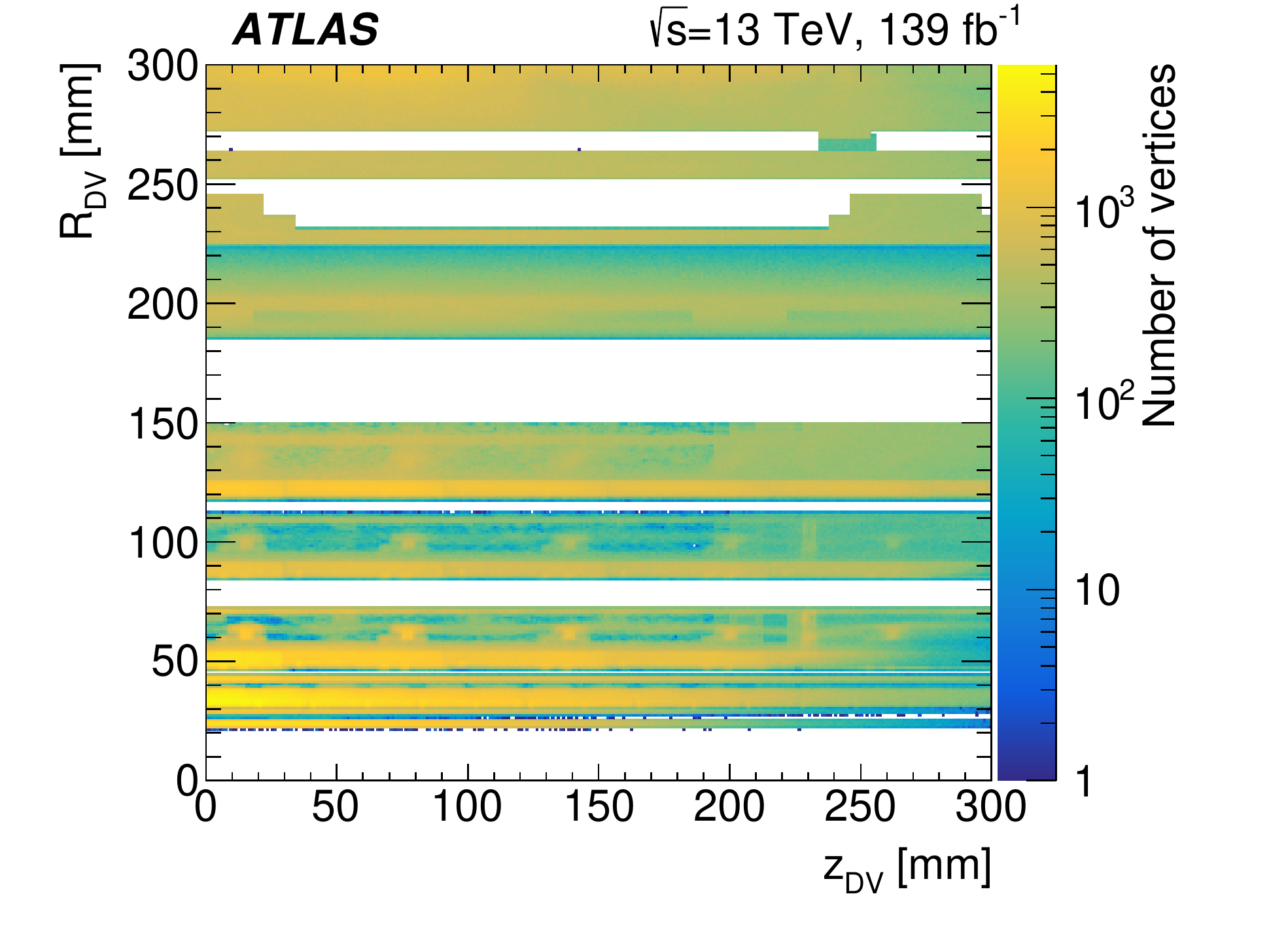}
\caption{}
\end{subfigure}
\caption{The positions in the (a) $x-y$ plane and (b) $r-z$ plane of reconstructed displaced vertices that are vetoed by the material map in the analyzed data sample. The right figure is restricted to positive $z$ values for presentation. The innermost area corresponds to the beampipe. This is surrounded by four pixel detector layers. The octagonal shape and outermost structures are due to support structures separating the pixel and SCT detectors.}
\label{fig:matmap}
\end{figure}
 
DVs passing the full selection must have at least five tracks, two of which must be selected tracks and a reconstructed invariant mass $\mDV$ greater than 10~\GeV. These requirements are designed to reduce the expected number of background events to around one. The $\mDV$ is calculated from the four-momenta of the tracks, assuming each track was produced by a charged pion. Only tracks satisfying specific quality selections, summarized in Table~\ref{tab:track_cleaning}, are used in the calculation of the DV track multiplicity and mass.  The selection criteria are designed to reject background DVs in the signal regions, while maintaining a high efficiency for signal vertices. As the background composition changes in different parts of the detector, these selections depend on $R_{\textrm{DV}}$, with different criteria before and after the beampipe ($R=25$~mm) and the last pixel layer ($R=145$~mm). Further requirements, aimed at rejecting wrong track-to-vertex associations, are applied on the azimuthal difference $\Delta\phi_{\textrm{PV--DV}}$ between the track direction and the vector connecting the primary vertex to the displaced vertex.
 
\begin{table}
\caption{Summary of the track cleaning selections used for signal DV candidates. The track $d_0$-significance is given by $|d_0|/\sigma(d_0)$.}
\label{tab:track_cleaning}
\begin{center}
\begin{tabular}{cl|ccc}
\hline\noalign{\smallskip}
& Region                         & $4 < R_{\textrm{DV}} < 25$~mm                                & $25 < R_{\textrm{DV}} < 145$~mm & $R_{\textrm{DV}} > 145$~mm \\
\noalign{\smallskip}\hline\noalign{\smallskip}
\multirow{4}{*}{Attached tracks} & Track \pT\ [GeV]               & $>2$                                                         & $>3^{\dagger}$                  & $>4$                       \\
& Track $d_0$-significance       & $>10$                                                        & $>15$                           & --                          \\
\noalign{\smallskip}\cline{2-5}\noalign{\smallskip}
& $\Delta\phi_{\textrm{PV--DV}}$ & \multicolumn{3}{c}{$<\pi/2$}                                                                                                \\
& Upstream veto                  & \multicolumn{3}{c}{No hits allowed with $R<R_{\textrm{DV}}$}                                                                \\
\noalign{\smallskip}\hline\noalign{\smallskip}
\multirow{3}{*}{Selected tracks} & Track \pT\ [GeV]               & $>2$                                                         & $>2^{\dagger}$                  & $>2$                       \\
& Track $d_0$-significance       & $>10$                                                        & --                               & $>10$                      \\
\noalign{\smallskip}\cline{2-5}\noalign{\smallskip}
& Upstream veto                  & \multicolumn{3}{c}{No hits allowed with $R<R_{\textrm{DV}}$}                                                                \\
\noalign{\smallskip}\hline\noalign{\smallskip}
\end{tabular} \\
$\dagger$~this requirement is tightened to $\pT>4$~\GeV\ for tracks with $\Delta\phi_{\textrm{PV--DV}}< 0.2$
\end{center}
\end{table}
 
This analysis uses two mutually exclusive signal regions, as defined in Table~\ref{tab:SRdef}. The High-\pT\ jet SR selects events that satisfy the High-\pT\ jet selection. The Trackless jet SR selects events which fail to satisfy the High-\pT\ jet selection but  do satisfy the Trackless jet selection. Both SRs require at least one DV passing the full selection.
 
\begin{table}
\caption{Summary of the signal region selections. The $x$ in the $n_{\text{jet/Trackless jet}}^{x}$ notation refers to the jet \pT~threshold in~\GeV. All jets are required to have $|\eta|<2.5$. The selection on $|\vec{R}_{DV} - \vec{R}_{CV}|$ considers all CVs in the event. }
\label{tab:SRdef}
\centering
\begin{tabular}{l|c|c}
\noalign{\smallskip}\hline\noalign{\smallskip}
Signal Region               & High-$\pt$ jet SR                                                                                           & Trackless jet SR                                                                 \\
\noalign{\smallskip}\hline\noalign{\smallskip}
& $n_{\text{jet}}^{250} \geq 4$ or $n_{\text{jet}}^{195} \geq 5$ or                                           & Fail High-\pT jet selection,                                                     \\
Jet selection               & $n_{\text{jet}}^{116} \geq 6$ or $n_{\text{jet}}^{90} \geq 7$                                               & $n_{\text{jet}}^{137} \geq 4$ or  $n_{\text{jet}}^{101} \geq 5$ or               \\
&                                                                                                             & $n_{\text{jet}}^{83} \geq 6$ or  $n_{\text{jet}}^{55} \geq 7$,                   \\
&                                                                                                             & $n_{\text{Trackless jet}}^{70} \geq 1$ or $n_{\text{Trackless jet}}^{50} \geq 2$ \\
\noalign{\smallskip}\hline\noalign{\smallskip}
& \multicolumn{2}{c}{$R_\text{DV}<300$~mm, $|z_\text{DV}|<300$~mm,}                                                                                                                              \\
DV preselection             & \multicolumn{2}{c}{ $\text{min}(|\vec{R}_\text{DV} - \vec{R}_\text{CV}|)>4$~mm, $\chi^2/n_\textrm{DoF}<5$,}                                                                                    \\
& \multicolumn{2}{c}{ $n_{\text{Selected tracks}}^{\text{DV}}\geq 2$, satisfy material map veto}                                                                                                    \\
\noalign{\smallskip}\hline\noalign{\smallskip}
$n^\text{DV}_\text{Tracks}$ & \multicolumn{2}{c}{$\geq 5$}                                                                                                                                                                   \\
$\mDV$                      & \multicolumn{2}{c}{$>$10~\GeV}                                                                                                                                                                 \\
\noalign{\smallskip}\hline\noalign{\smallskip}
\end{tabular}
\end{table}
 
The total efficiency of the high-$\pt$ jet SR for selecting generated events in the EWK RPV model ranges from 20\% to 70\% for ${m(\ninoone)=500}$ \GeV\ to ${m(\ninoone)=1.7}$ \TeV\ and $\tau(\ninoone)=0.1$~ns. For the Strong RPV model, the total efficiency of the high-$\pt$ jet SR ranges from 50\% to 80\% for ${m(\ninoone)=200}$ \GeV\ to ${m(\ninoone)=1250}$ \GeV\ with $m(\gluino)=2.6$ \TeV\ and $\tau(\ninoone)=0.1$~ns. The total efficiency of the Trackless jet SR ranges from 7\% to 40\% for the EWK RPV model with $m(\ninoone)=300$ \GeV\ to $m(\ninoone)=700$ \GeV\ and $\tau(\ninoone)=0.1$~ns, and does not have significant sensitivity to the Strong RPV model. Selection efficiency of both of the signal regions decreases for samples with neutralinos with shorter or longer lifetimes, due largely to the fiducial volume acceptance.

\FloatBarrier
\section{Backgrounds}
 
There are no high-mass SM particles which produce displaced decays which satisfy the selections required in this search. Searches for displaced, multi-track vertices benefit from a small background which is caused predominantly by low-mass vertices being merged together and misclassified as a high-mass vertex.
The backgrounds arise from instrumental and algorithmic effects and can be described by three main sources: \HI, \AX, and \MV.
 
Hadronic interactions arise from particles interacting with the nuclei of detector material while traveling through the detector. If multiple charged particles are produced in this interaction, then they may be reconstructed as a displaced vertex. Due to the kinematics of this process, the charged particles resulting from the interaction are typically collimated, giving rise to relatively low-mass displaced vertices. Rare interactions occurring in less collimated decay products may result in a high-mass displaced vertex. The bulk of these DVs are localized in the material-dense regions of the detector and are removed by the material map veto. The residual background is due to hadronic interactions in the less dense regions of the detector or those that evade the material map veto due to the resolution of the vertex position or due to inefficiencies in the material map. This background class also includes vertices created by regular long-lived SM particles decaying in their natural decay mode. As the mass of long-lived particles in the SM are light ($\lessapprox5$ \GeV), these decays typically result in low-mass reconstructed vertices.
 
Low-mass displaced vertices that are crossed by an unrelated track are referred to as accidental crossings. A low-mass DV may be misclassified as a high-mass DV with higher track multiplicity if an unrelated track passes sufficiently close to the vertex to be attached to it during reconstruction. These tracks frequently form a large angle with the other tracks in the low-mass SM vertices they cross, resulting in a large increase in the invariant mass of the vertex.
 
The last category, merged vertices, considers the cases when two or more low-mass displaced vertices occur in the vicinity of each other. In these cases, the vertexing algorithm may mistakenly include them in the same vertex. This may increase the mass of the displaced vertex to the signal region. Vertices from decays of long-lived SM particles, hadronic interactions, or accidental crossings may be merged by the vertexing algorithm, resulting in a high-mass DV that satisfies the signal region selection.
 
Backgrounds are estimated by using an inclusive data-driven technique that predicts the rate of DVs from all three sources above. This method is described in Section~\ref{sec:inclusivebackground} and validated in Section~\ref{sec:validationbackground}. Additionally, an alternative estimate is presented in Section~\ref{subsec:alternativebkg} that is used to predict the three different sources of backgrounds separately and produce estimates for their distribution in the reconstructed mass of the DV. This method is used as additional validation and its results are compatible with the inclusive method. Furthermore, the alternative estimate is used to evaluate a systematic uncertainty on the inclusive method as described in Section~\ref{sec:uncertainties}.
 
\subsection{Inclusive background estimation}
\label{sec:inclusivebackground}
 
The background estimation used in this analysis relies on the assumption that in a multijet final state, DV production is correlated with the presence of jets. All three sources of background DVs are correlated with track density, and the track density is correlated with the presence of jets in the event. Track-jets are chosen to maximize this correlation.
 
The method relies on the measurement of the probability that a SR-like DV is produced in proximity of a track jet in a statistically independent event selection from the SR. The probability is measured in a control region of data and is split into two parts. The first part, \dvprob, is parameterized as a function of track jet properties and the second part, an extrapolation factor $f$, is defined inclusively. The extrapolation factor takes into account differences in the DV mass and track multiplicity requirements in the CR and in the SR. The \dvprob\ is used to assign a weight to track-jets in events passing the SR jet selections. The sum of the weighted track-jets, multiplied by the extrapolation factor $f$, is taken as the total number of events with a background DV expected in the SR.
 
To study the correlation between DVs and track-jets in data, a control region is designed which contains track-jets but which also limits the contamination from signal models of interest. The CR is populated with data events collected using a single photon trigger, requiring a photon with transverse energy $\ET>140$ \GeV. This trigger is selected because it does not alter the flavour composition of jets in the CR relative to those in the signal region. Events in the CR must fail to satisfy the jet selection requirements of both signal regions. Finally, events in the CR must have at least three track-jets with $\pT>20$ \GeV. There is negligible signal contamination in this CR from the two simplified models considered in this search.
 
The jet-DV probability is parameterized as a function of both the track jet \pT and track jet track multiplicity $N_\text{trk}^\text{jet}$ to take into account residual differences between the CR and the SRs. The fraction of events in the CR which contain a preselected DV with $\mDV>3$~GeV and $\ntrk>2$ as a function of track jet multiplicity is shown in Figure~\ref{fig::DVsInCRa}. Events with more track-jets are found to be more likely to contain a DV.
The proximity of a DV and a track jet is calculated with $\Delta R_{\mathrm{DV,jet}} = \sqrt{\Delta\eta^2+\Delta\phi^2}$ where $\Delta\eta$ and $\Delta\phi$ are measured between the PV--DV direction and the jet axis. The average $\Delta R_{\mathrm{DV,jet}}$ as a function of the DV track multiplicity is shown in Figure~\ref{fig::DVsInCRb}. DVs with a higher track multiplicity tend to be produced in closer proximity to track-jets, which is a  sign that DVs which are more SR-like ($\nTrk\geq5$) have a stronger correlation to track jet activity.
 
\begin{figure}[ht!]
\centering
\begin{subfigure}[b]{0.49\textwidth}
\includegraphics[width=\textwidth]{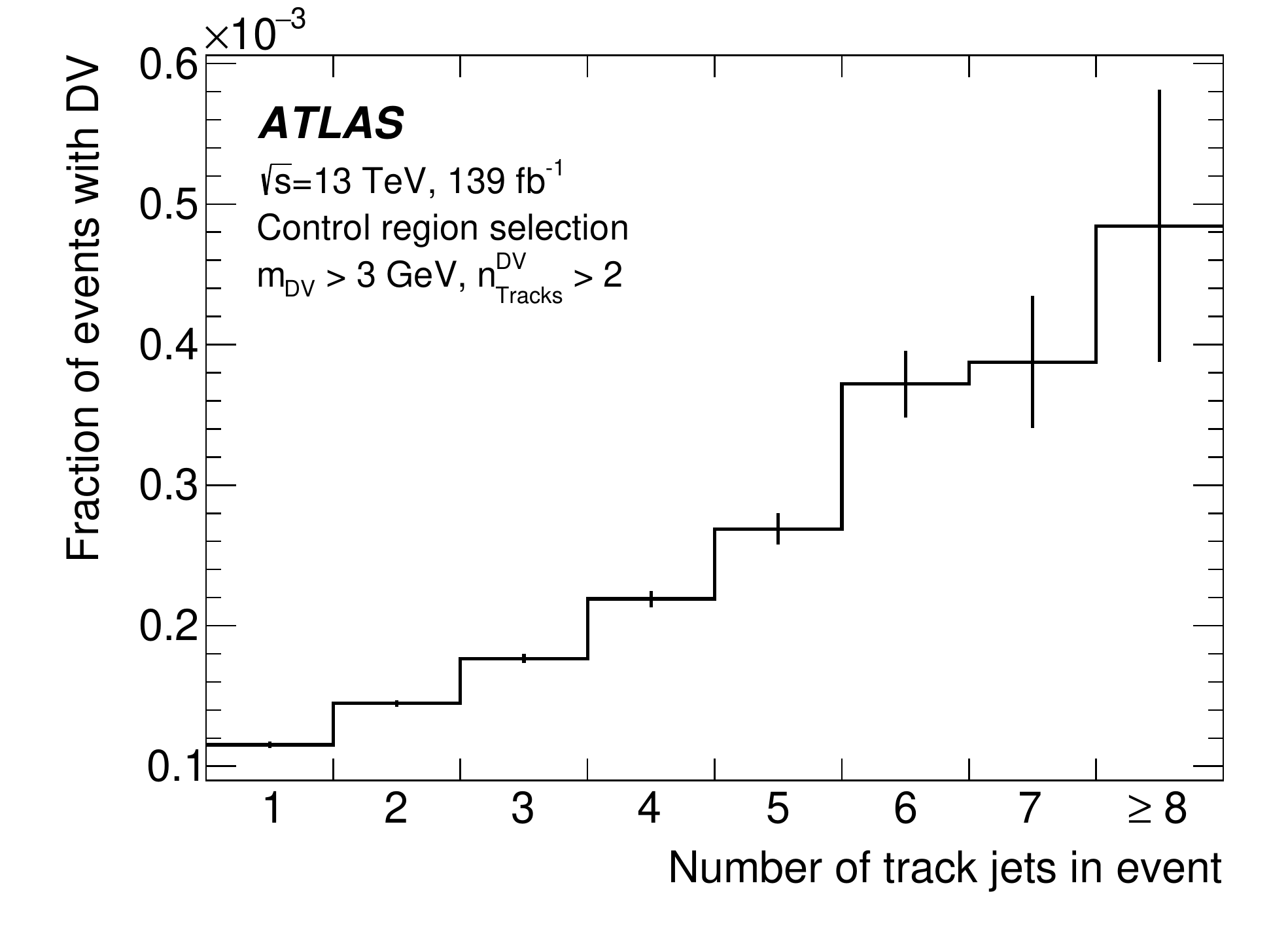}
\caption{}
\label{fig::DVsInCRa}
\end{subfigure}
\begin{subfigure}[b]{0.49\textwidth}
\includegraphics[width=\textwidth]{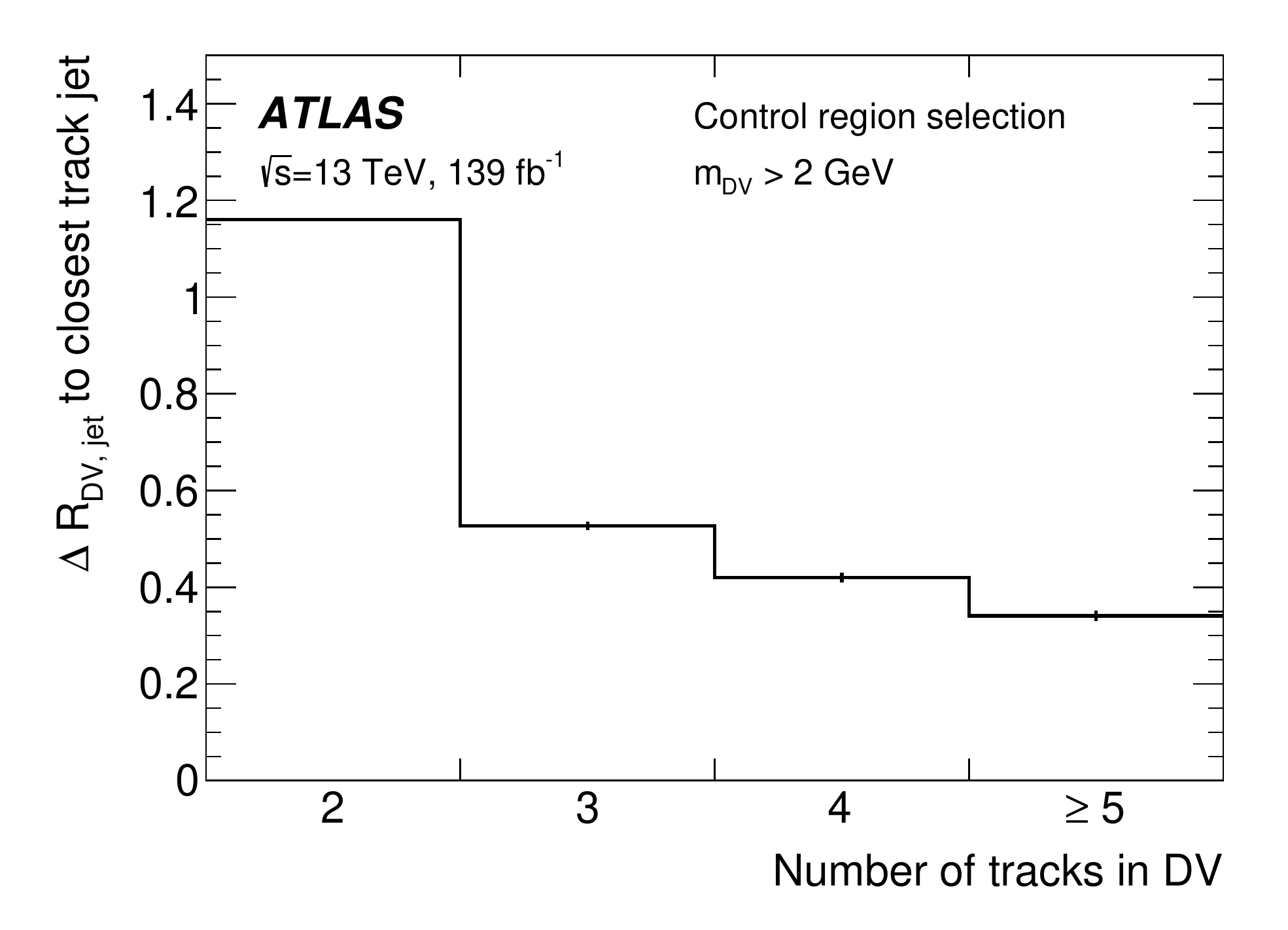}
\caption{}
\label{fig::DVsInCRb}
\end{subfigure}
\caption{In (a), the fraction of events in the CR with a reconstructed DV satisfying $\mDV > 3$ \GeV\ and $\nTracks> 2$ is shown. In (b), the average distance of DVs with $\mDV > 2$ \GeV\ to the closest track jet in the event is shown, as a function of $\nTrkCaption$.}
\end{figure}
 
The SR-like DVs in the CR are matched to the closest track jet in the event using $\Delta R_{\mathrm{DV,jet}}$. Each DV is matched to only one track jet in each event, and the DV position is not required to be within the jet. More than $70\%$ of DVs are within the $\Delta R=0.4$ cone of the nearest track jet.
The $\Delta R_{\mathrm{DV,jet}}$ distribution does not depend strongly on $\mDV$, but shifts to lower values as $\nTracks$ increases.
 
The jet-DV probability is computed as the ratio of the number of track-jets with a matched DV over the total number of track-jets in the CR. 
All DVs in the CR satisfying $\nTracks>3$ and $\mDV>5$ \GeV\ are considered for the determination of \dvprob. The DV mass and track multiplicity requirements in the CR are looser than the SR-DV requirements so that there is adequate statistical precision in the CR to calculate the jet-DV probability. The jet-DV probability is found to range between $6 \times 10^{-7}$ and $2 \times 10^{-6}$, with a mild dependence on the track jet \pT and $N_\text{trk}^\text{jet}$.
 
In events passing the High-$\pT$ (Trackless) jet selections, each track-jet is assigned a weight which is equal to the corresponding \dvprob, in order to estimate the number of background DVs expected in the High-$\pT$ (Trackless) jet SR. The weighted sum is then multiplied by an additional extrapolation factor $f$ which considers the difference between the DV mass and track multiplicity requirements used to calculate the jet-DV probability and in the SR selection. The extrapolation $f$ is calculated by taking the ratio of the number of events with a reconstructed DV satisfying $\nTracks>4$ and $\mDV>10$ \GeV\ relative to the number of events with a reconstructed DV satisfying $\nTracks>3$ and $\mDV>5$ \GeV\ in the CR. The extrapolation $f$ was determined to be $0.038\pm0.022$, where the uncertainty is propagated from the statistical variation of the number of events in the CR.
 
The method predicts	respectively $0.46^{+0.27}_{-0.30}$ background events in the High-$\pt$ jet SR and $0.83^{+0.51}_{-0.53}$ background events in the Trackless jet SR.
 
\FloatBarrier
\subsection{Validation of background model}
\label{sec:validationbackground}
 
Several statistically independent validation regions are defined and used to test the predictions of the background estimation method.
Validation regions are defined in the SR sidebands by selecting events which pass the full signal region selection, except they contain at least one DV with $\mDV<10$ \GeV\ or $\nTracks<5$. A second set of validation regions is defined with events which pass all SR selections, but contain a high-mass DV that fails to meet the criteria of the material veto. For each validation region, a separate jet-DV probability is calculated from events in the CR. There are few events in the CR which satisfy the same DV selection criteria as the SR sideband validation regions with $\nTracks=4$, $\mDV>10$ \GeV\ and with $\nTracks\geq5$, $5<\mDV<10$ \GeV. Therefore, the $\mDV$ or $\nTracks$ requirements are loosened in order to calculate the jet-DV probability in a similar manner as the procedure performed for the estimate of background in the SR. The agreement between the background estimate and observed event counts in each of the validation regions is shown in Figure~\ref{fig::vrsummary}.
 
\begin{figure}[ht!]
\centering
\begin{subfigure}[b]{0.49\textwidth}
\includegraphics[width=\textwidth]{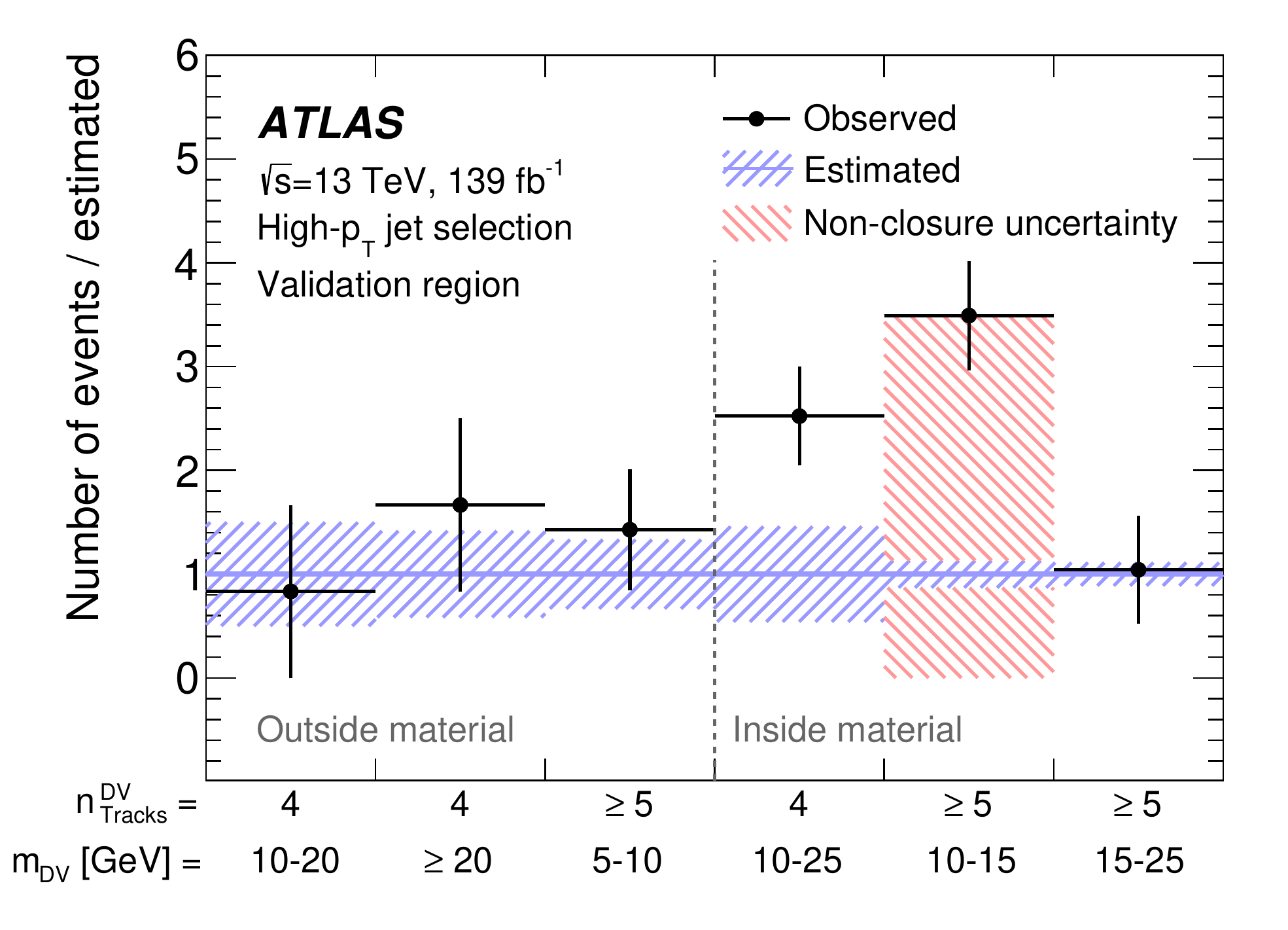}
\caption{}
\label{fig::vrsummary_highpt}
\end{subfigure}
\begin{subfigure}[b]{0.49\textwidth}
\includegraphics[width=\textwidth]{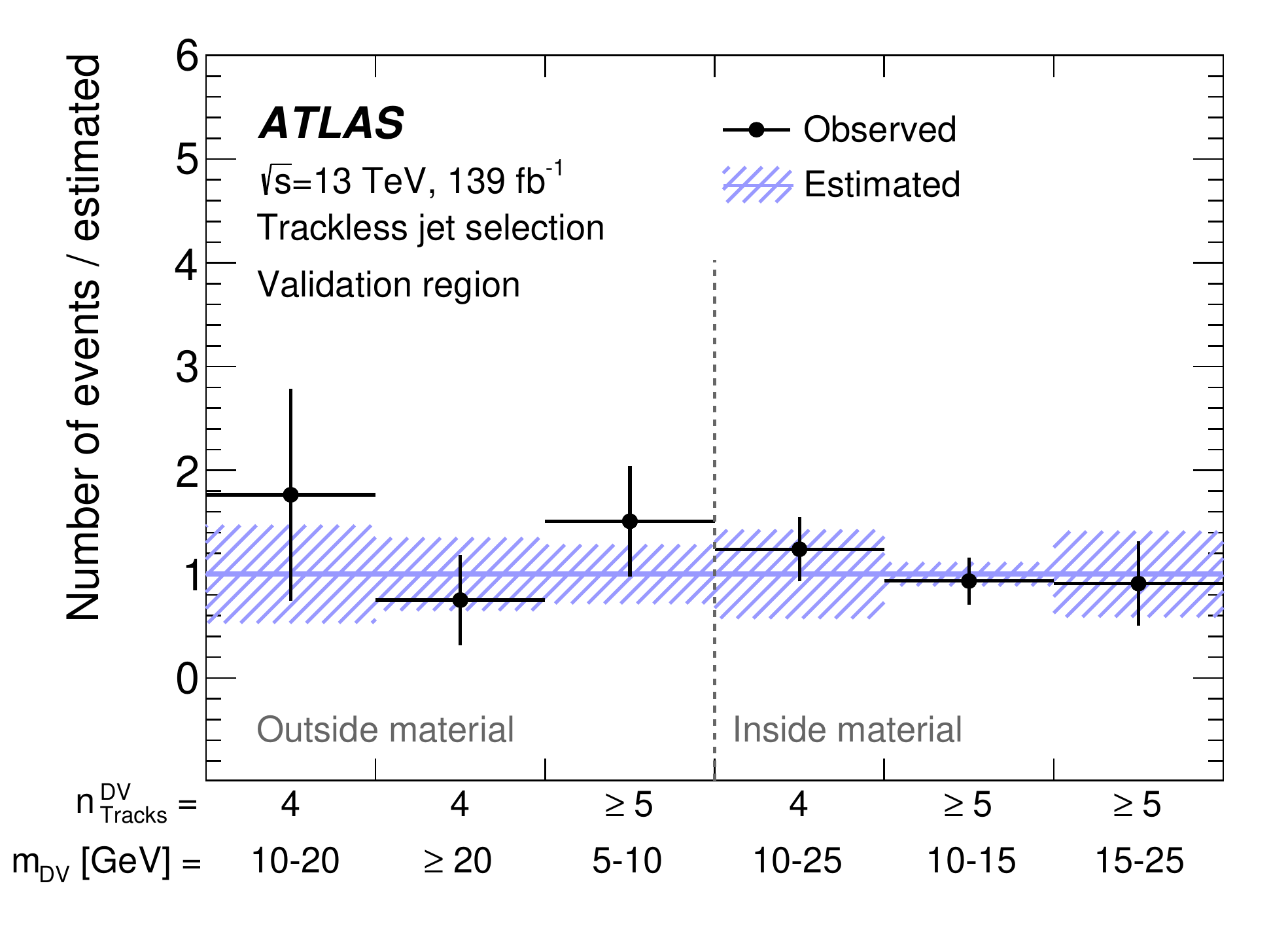}
\caption{}
\end{subfigure}
\caption{Background estimates relative to expectations for (a)  the High-$\pt$ jet selection and (b) the Trackless jet selection showing validation regions with varying $n_{\text{Tracks}}^{\text{DV}}$ and $\mDV$ requirements. Validation regions labeled \textit{Outside (Inside) material} contain events with a displaced vertex which is accepted (rejected) by the material map veto requirement. The error band dashed with a positive slope includes all systematic uncertainties. The size of the non-closure uncertainty is shown by the band dashed with a negative slope, which is not propagated to the other validation regions in this figure.}
\label{fig::vrsummary}
\end{figure}
 
Within uncertainties, good agreement is found between the background model and data in all of the SR sideband validation regions, as well as the inside material VRs for the Trackless jet selection. For events satisfying the High-$\pt$ jet selection criteria  inside of the  material validation regions, there is a significant underestimation of the background in the $\nTracks=4$, ${10<\mDV<25}$ \GeV\ and $\nTracks\geq5$, ${10<\mDV<15}$ \GeV\ regions. These regions are significantly populated by hadronic interactions, which are expected to have a minimal contribution to the background in the signal region. The difference between the observed yields and the expectation from the background prediction in the $\nTracks\geq5$, $10<\mDV<15$ \GeV\ region is used to define a symmetric uncertainty for this background component, as shown in Figure~\ref{fig::vrsummary_highpt} and detailed in Section~\ref{sec:uncertainties}.
 
\subsection{Alternative background estimate}
\label{subsec:alternativebkg}
 
Further validation of the estimate is performed with an alternative approach that aims to estimate each source of background independently. Data-driven techniques are used to build mass templates which describe the shape of $\mDV$ from each specific background source, which are then normalized and combined into a complete background estimate. Though not used directly in the interpretation of the search results, the alternative background estimate serves as an important validation of the background composition and the robustness of the inclusive background estimate.
 
Mass templates of merged vertices with varying track multiplicities are built by randomly merging pairs of DVs taken from different events in data. Only DVs with a distance significance of $S<10$ are merged, where $S$ is defined as the distance between vertices divided by its uncertainty. The mass templates are normalized by comparing the significance of the distance between pairs of DVs in the same event and in distinct events, as shown in Figure~\ref{fig::mvdist}. The distribution of the distance significance taken from distinct events is normalized to match the distance significance distribution of DV pairs from the same event in the $S>10$ region, where merging cannot occur. The deficit of DV pairs in the same event relative to DV pairs taken from distinct events in the $S<10$ region is set as the overall normalization of the merged vertex mass template. In the assumption that the merging probability is small, this normalization procedure ensures that the difference of these two distributions is equal to the number of merged vertices expected~\cite{STDM-2015-02}. To increase the number of available events for the determination of the normalization factor, only the cleaning requirements on track $d_0$-significance and the track $\pt$ for tracks with $\Delta\phi_{\textrm{PV--DV}}< 0.2$ are applied. All other requirements are removed, as the quantities used in the selection criteria were found to have no correlation with the merging rate. The mass templates are built with the same loosened DV-track cleaning requirements, and the full DV-track cleaning is applied to the mass templates after they are normalized.
 
This background estimation method is sensitive to the presence of correlated DVs in the same event. Pairs of DVs are correlated if they originate from the same cascade hadronic interaction, for example. DVs which are correlated tend to be reconstructed close together with a small $\Delta\phi$, leading to an excess of DV pairs from the same event with a small distance significance between them. To evaluate the impact that correlated DVs have on this estimate, two sets of merged vertex mass templates for each track multiplicity of DVs are produced. The first set of templates is produced without a $\Delta\phi>0.2$ selection, while the second set of templates is produced with the $\Delta\phi>0.2$ selection. The nominal merged vertex background estimate is derived from the bin-by-bin average of the two sets of templates.
 
\begin{figure}[b!]
\centering
\includegraphics[width=0.6\textwidth]{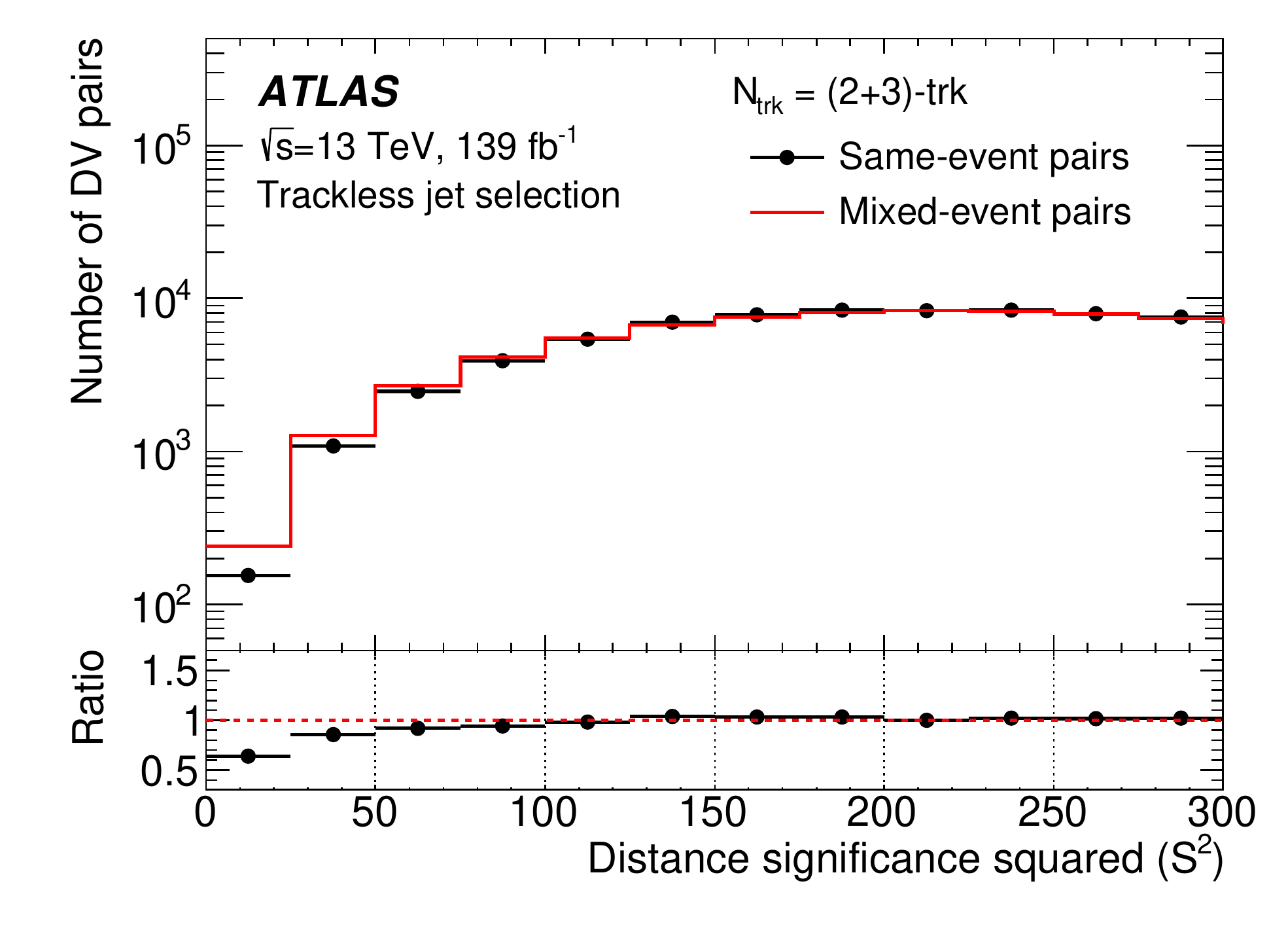}
\caption{Squared significance of the distance between pairs of preselected DVs with two and three tracks (denoted by $\nTracks = (2+3)$-trk) taken from the same event and from distinct events passing the Trackless jet selection. The mixed event distribution is normalized to match the same event distribution in the $S^2>100$ region.}
\label{fig::mvdist}
\end{figure}
 
The contribution from high-mass DVs with an accidental-crossing track is estimated as follows. Templates for ($n$+1)-track vertices are constructed by adding an accidental-crossing track to $n$-track vertices from data. The accidental-crossing track is drawn randomly from a collection of tracks reconstructed in data which have been incorrectly attached to a DV. The tracks are identified using displaced $\Kshort\rightarrow \pi^{+}\pi^{-}$ decays in data. To identify accidental-crossing tracks, 3-track DVs where two of the tracks can be combined to form a vertex consistent with the $\Kshort$ mass are identified. The third track in the vertex is added to the collection if it passes the DV-track cleaning. The track's $\pt$, $\eta$, and relative azimuthal angle to the direction of the PV-DV vector ($\Delta\phi$) are added to the collection. The ($n$+1)-track DV mass distribution is built by recalculating the mass of $n$-track DVs in data after attaching one additional track to it.
 
The behaviour of the accidental-crossings background is expected to differ depending on the position of the DV in the fiducial volume. This is due to multiple factors, for example the vertexing reconstruction efficiency and the larger incident angles of tracks crossing DVs at large radii. Therefore this background estimate is evaluated separately in four radial detector regions\footnote{The boundaries for these radial regions are at $R= 0$, 38, 120 and 300~mm.}, divided approximately by material structures in the inner detector volume within the fiducial region.
 
The mass templates for the accidental-crossing background are normalized with a \textit{crossing factor}, which is a measure of the probability that a DV is reconstructed with an unrelated track. The crossing factor is calculated by measuring the fraction of all $\Kshort\rightarrow \pi^{+}\pi^{-}$ decays reconstructed in data which have an accidentally crossing track. The number of $\Kshort$ decays reconstructed without (with) an accidental crossing track is calculated using the mass spectrum of 2-track DVs (2-track combinations in 3-track DVs), shown in Figure~\ref{fig::ksmass}. The mass spectrum is fit to a Gaussian distribution peak and a linear background. The linear background is subtracted from the mass spectrum and the number of entries in the mass spectrum within 40~\MeV\ of the $\Kshort$ mass gives the number of reconstructed $\Kshort\rightarrow \pi^{+}\pi^{-}$ decays. The level of non-$\Kshort$ background in the $\Kshort$ mass window is less than 10\% in the 2-track DV mass spectra and ranges from 20--70\% in the 3-track DV mass spectra, depending on the radial region.
\begin{figure}[ht!]
\centering
\begin{subfigure}[b]{0.49\textwidth}
\includegraphics[width=\textwidth]{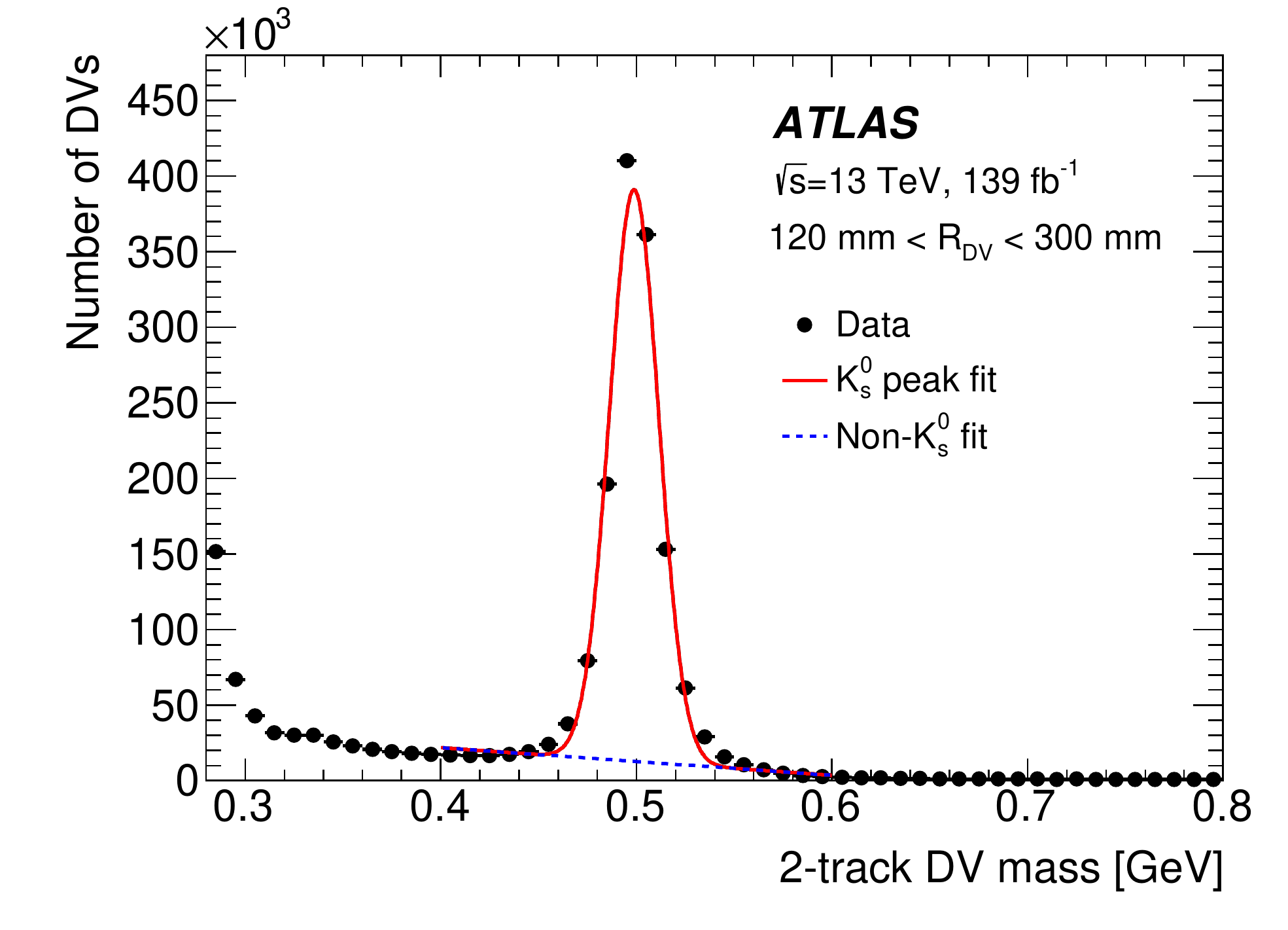}
\caption{}
\end{subfigure}
\begin{subfigure}[b]{0.49\textwidth}
\includegraphics[width=\textwidth]{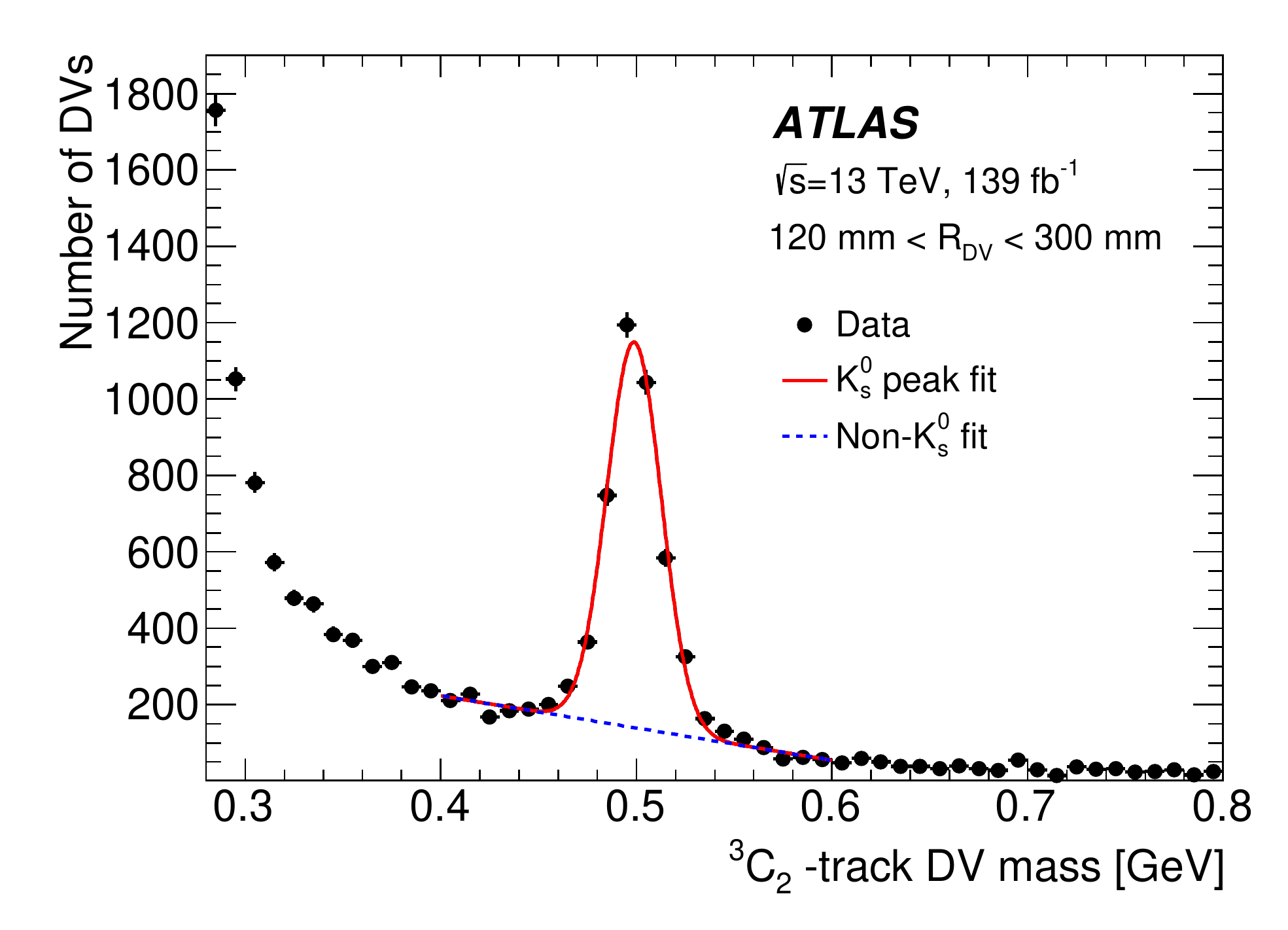}
\caption{}
\end{subfigure}
\caption{The mass spectrum of preselected (a) 2-track DVs and (b) 2-track combinations of 3-track DVs ($^3\text{C}_2$) in one radial region. The mass spectrum is fit to a Gaussian function centered on the $\Kshort$ mass and a linear background. The data in the $\Kshort$ mass spectrum after subtracting the fitted background is used to compute the $\Kshort$ crossing factors.}
\label{fig::ksmass}
\end{figure}
 
In each radial region, two crossing factors are calculated. The first is dedicated to accidental crossings of selected tracks, which is required to estimate the contribution of $n$-track vertices with only one selected track that would otherwise fail to satisfy the DV selection criteria. The second considers all tracks and is used to estimate the contribution of $n$-track vertices with two or more selected tracks. In order to increase the number of reconstructed $\Kshort$ decays available for these calculations, the innermost two radial regions are merged and a single set of crossing factors are calculated for both regions. The crossing factors, shown in Figure~\ref{fig:crossingfactor}, range from $8.26\times10^{-5}$ in the innermost radial region to $2.91\times10^{-3}$ in the outermost radial region. The crossing factor increases at larger radii due to the decreasing vertex position resolution.
 
\begin{figure}[b!]
\centering
\includegraphics[width=0.6\textwidth]{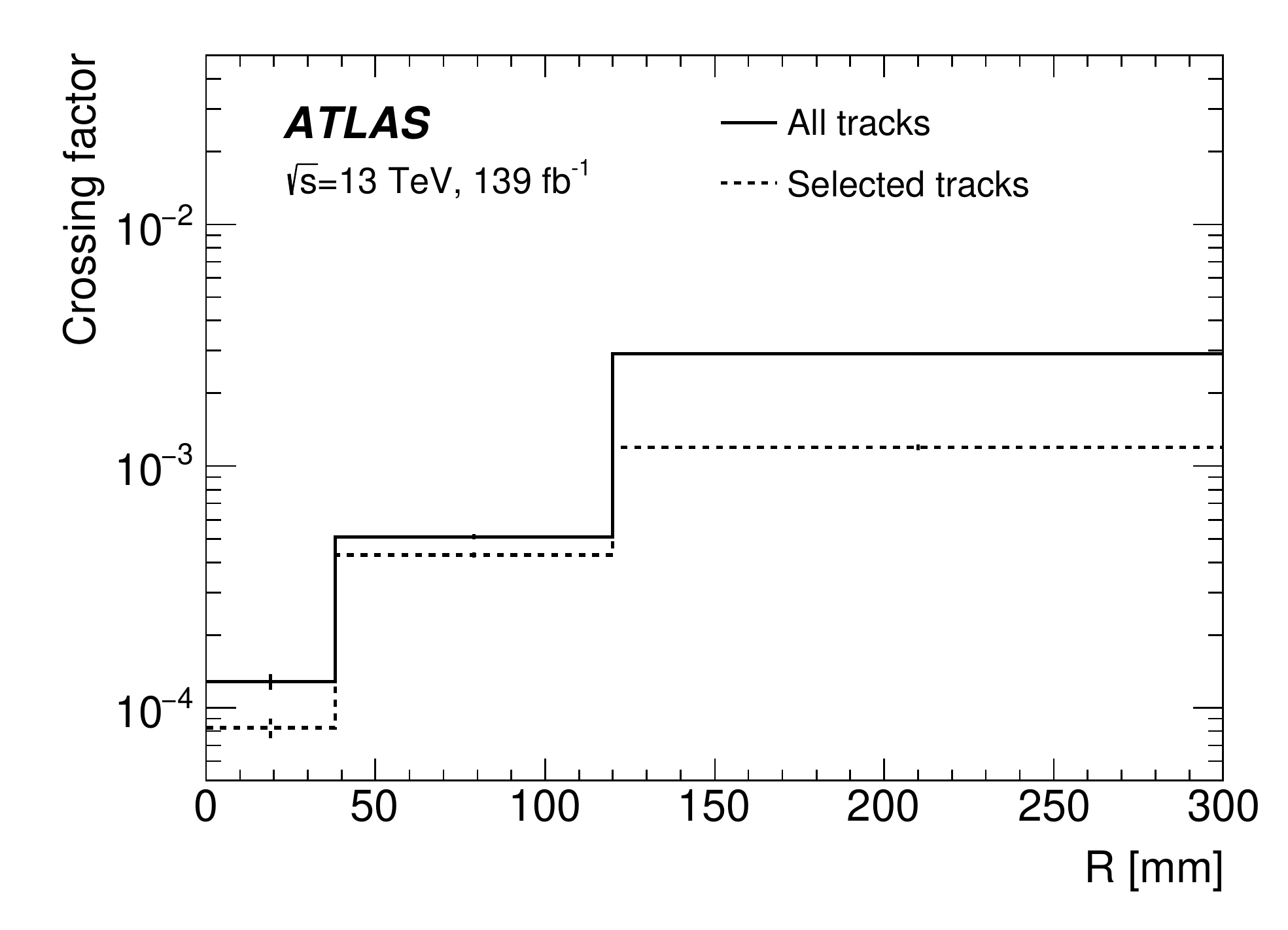}
\caption{Crossing factors computed in each radial region. The bars show the statistical uncertainties.}
\label{fig:crossingfactor}
\end{figure}

The component of background arising from hadronic interactions dominates in the low-mass region of the $\mDV$ distribution. Therefore, to estimate this background in the signal region, the $\mDV$ distribution is fit with an empirical function modelling the sharp increase below 2~\GeV\ and follows an exponentially falling distribution in the high-mass region. The function is fitted to data in the $\mDV<10$ \GeV\ region and extrapolated to the SRs with $\mDV>10$ \GeV. Studies based on simulated events indicate that the contamination from background DVs originating from sources other than hadronic interactions in the $\mDV<10$ \GeV\ region is expected to be negligible. There is a secondary component of hadronic interactions present in the high-$\mDV$ region with a shallower exponential tail caused by the presence of elastic hadronic interactions, where the mass of the DV is dominated by a single track with small $\Delta\phi_{\textrm{PV--DV}}$. The difference between these exponential distribution tails is measured in simulated multijet events and used to apply a correction to the hadronic interactions template derived from data.

The normalized background estimates for the individual components are added to form a total expected background. The expected and observed event counts in the $\nTracks=4$ sideband validation regions are shown in Figure~\ref{fig:combinedvalidation}. Good agreement is observed across the entire mass range.
 
\begin{figure}[h]
\centering
\begin{subfigure}[b]{0.48\textwidth}
\includegraphics[width=\textwidth]{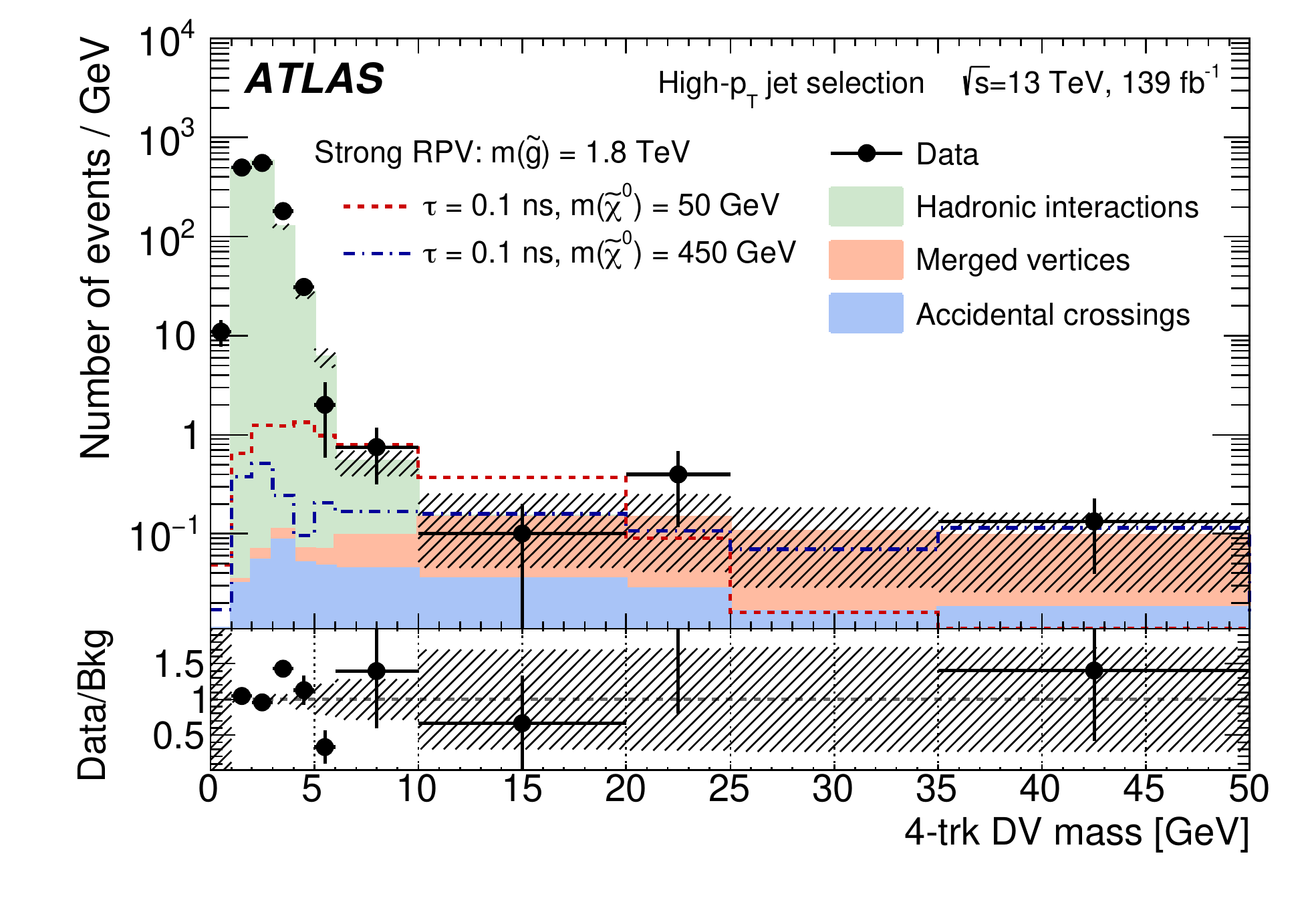}
\caption{}
\end{subfigure}
\begin{subfigure}[b]{0.48\textwidth}
\includegraphics[width=\textwidth]{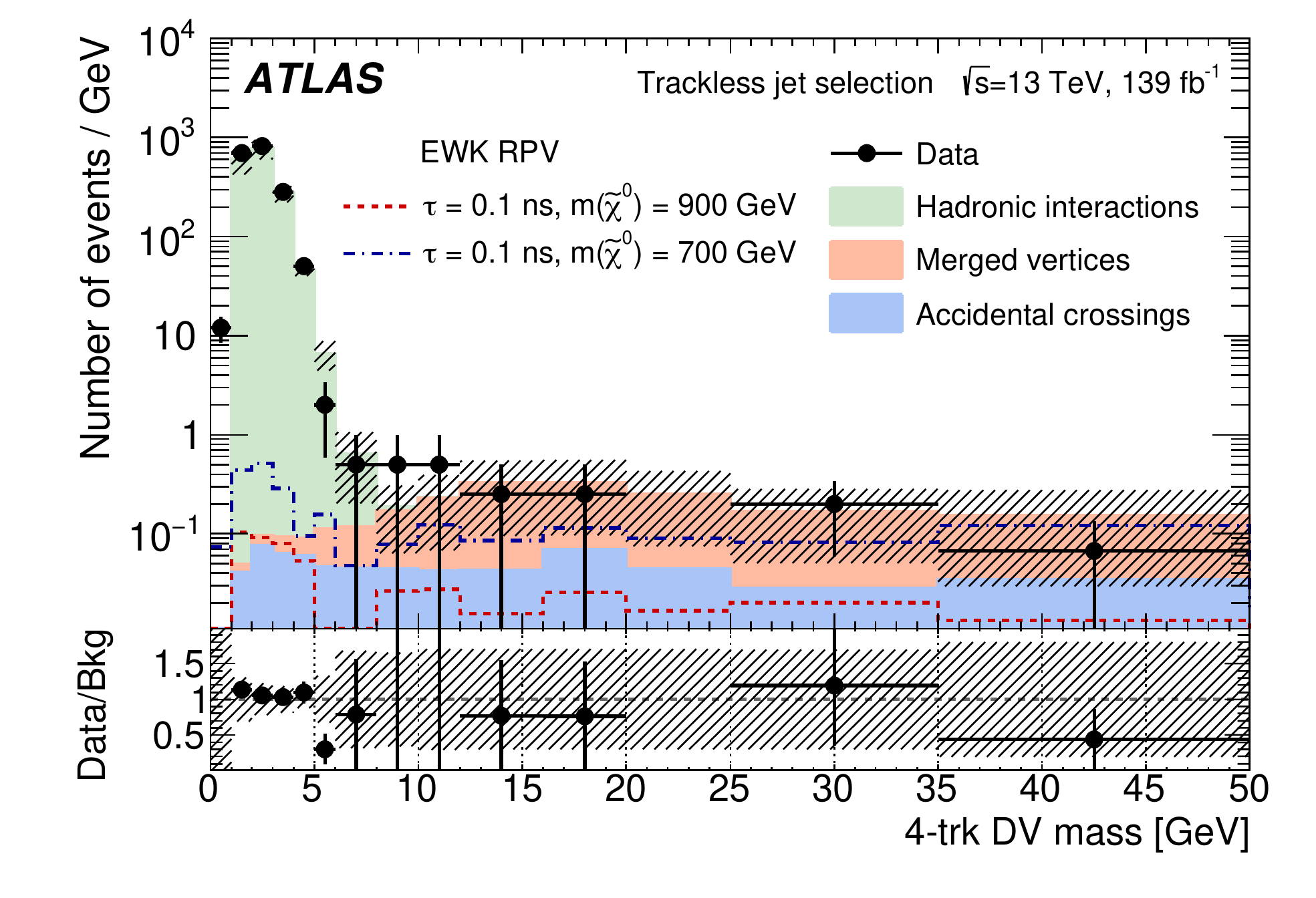}
\caption{}
\end{subfigure}
\caption{The combined background predictions in events passing the (a) High-$\pt$ jet selection and (b) Trackless jet selection 4-track DV validation regions. The overflow is included in the rightmost bin.}
\label{fig:combinedvalidation}
\end{figure}
 
The total expected background in the SR is calculated by integrating all $\geq5$-track mass templates in the $\mDV>10$ \GeV\ region, shown for the High-$\pt$ jet and Trackless jet SRs in Table~\ref{tab:finalestimates}. The inclusive and component-based estimates are consistent within uncertainties, thereby providing a successful cross-check of the inclusive background estimation technique. The inclusive method is chosen as the main method because it gives a smaller uncertainty and it has a more robust validation technique.
\begin{table}
\caption[Combined background estimates in the signal regions]{The separate background estimates in the signal regions. The systematic uncertainties shown for each estimate are described in Section~\ref{subsec:combunc}. The uncertainties on the combined background estimate are calculated assuming no correlation between the separate estimates. The rightmost column contains the background estimates from the inclusive method, described in Section~\ref{sec:inclusivebackground}.}
\label{tab:finalestimates}
\centering
\begin{tabular}{l|ccc|c|c}
\noalign{\smallskip}\hline\noalign{\smallskip}
\multirow{2}{*}{Region} & Merged        & Hadronic        & Accidental    & \multirow{2}{*}{Combined} & \multirow{2}{*}{\hspace{0.8mm}Inclusive\hspace{0.8mm}} \\
& vertices      & interactions    & crossings     &                           &                                                        \\
\noalign{\smallskip}\hline\noalign{\smallskip}
High-$\pt$ jet SR       & $0.79\pm0.66$ & $0.006\pm0.018$ & $0.28\pm0.21$ & $1.08\pm0.69$             & $0.46^{+0.27}_{-0.30}$                                 \\[5pt]
Trackless jet SR        & $1.5\pm1.1$   & $0.248\pm0.077$ & $0.32\pm0.24$ & $2.1\pm1.1$               & $0.83^{+0.51}_{-0.53}$                                 \\
\noalign{\smallskip}\hline\noalign{\smallskip}
\end{tabular}
\end{table}

\FloatBarrier
\section{Uncertainties}
\label{sec:uncertainties}
 
In the following sections  uncertainties derived for the inclusive background estimation, the alternative background estimate, and the expected signal yield are described.
 
\subsection{Uncertainties in the inclusive background estimation}
\label{subsec:inclunc}
The systematic uncertainties associated with the background estimate are evaluated for each SR selection.
Four sources of uncertainty are taken into account. First, the statistical uncertainty in the number of events with a SR-like DV in the CR is propagated to the final estimate. A toy method is used which performs 10,000 pseudo-experiments. In each pseudo-experiment, the number of events in the CR with a SR-like DV is varied bin-by-bin within its statistical uncertainty and used to calculate a new jet-DV probability distribution and extrapolation factor $f$. The standard deviation of the background estimate from all statistically varied toy experiments is taken as an uncertainty in the final background estimate.
 
The uncertainty in the residual dependence of the jet-DV probability on the amount of jet activity in the event is estimated by computing a new jet-DV probability which is parameterized as a function of the number of track-jets in the event. No statistically significant dependence of the jet-DV probability as a function of the jet activity in the event was found. However, to estimate any residual dependence below the statistical accuracy, a new jet-DV probability is computed which is parameterized as a linear function of the number of track-jets in the event. The fitted parameters in the linear function are varied within their uncertainties and the resulting jet-DV probability is used to estimate the background in the signal region. The largest variation in the background estimates is taken as an uncertainty.
 
A third uncertainty is applied to account for the non-closure of the estimate in some of the validation regions. This method underestimates the number of DVs reconstructed inside of material in events which satisfy the High-\pT\ jet selection criteria, as shown in Figure~\ref{fig::vrsummary_highpt}. Studies of simulated background events and the alternative background estimate indicate that the validation region in which this non-closure is the largest (events containing a DV with $\nTracks\geq5$ and ${10~\text{GeV}<\mDV<15}$ \GeV\ reconstructed inside of material) are populated dominantly ($>99$\%) with background DVs from hadronic interactions. The same methods indicate that the expected contribution of hadronic interactions in the High-$\pT$ jet SR is very small ($<1$\%), due largely to the material map veto and $\mDV$ requirements. The ratio of observed to expected events in the validation region with the largest non-closure is assigned as an uncertainty in the hadronic interactions background component. To determine an uncertainty in the inclusive background in other regions, the uncertainty in the hadronic interactions background component is scaled by the expected contribution of hadronic interactions to the total background in each region. The expected contribution of hadronic interactions in each region is calculated using the alternative background estimate. This uncertainty covers the remaining discrepancies between observed and expected background in the validation regions shown in Figure~\ref{fig::vrsummary_highpt}. The resulting uncertainty assigned to the inclusive background in the High-$\pT$ jet SR is 4\%.
 
Lastly an uncertainty is assigned to account for differences in the pile-up distributions between events in the CR and events which pass the jet selections of the SR. Events in the CR are reweighted so that their pile-up distribution matches that of events passing the SR jet selections and the background estimate is re-evaluated. The difference between the background estimate with and without pile-up-reweighting is used to define a correction factor, which is applied to both the nominal background estimate and its uncertainties. Furthermore the magnitude of this correction factor is applied as an additional uncertainty to the nominal background estimate.
 
These uncertainties are assumed to be uncorrelated, and added in quadrature to estimate the total systematic uncertainty. The final estimate of background in the signal regions with their associated uncertainties is listed in Table~\ref{tab::finalsrestimate}.
 
\begin{table}[!ht]
\caption[]{Systematic uncertainties associated with the background estimate. The ``CR statistical'' uncertainty is the propagation of the statistical variations of the CR to the estimate, while the ``Non-linearity'' uncertainty refers to the residual dependence of the jet-DV probability on the amount of jet activity in the event. The ``VR non-closure'' is calculated from the non-closure of the estimate inside of material validation regions. The ``Pile-up" uncertainty is calculated by reweighting events in the CR to match the pileup distribution of events in each SR.}
\label{tab::finalsrestimate}
\begin{center}
\begin{tabular}{c|cccc}
\noalign{\smallskip}\hline\noalign{\smallskip}
Region           & \multicolumn{4}{c}{Uncertainty source}                                                       \\
& CR statistical                         & Non-linearity       & VR non-closure & Pile-up       \\
\noalign{\smallskip}\hline\noalign{\smallskip}
High-\pT jet SR  & $\pm\ 58\%$                            & $^{+10\%}_{-27\%}$  & $\pm\ 4\%$     & $\pm\ 5.9\%$ \\  \\[-1em]
Trackless jet SR & $\pm\ 58\%$                            & $^{+6.3\%}_{-17\%}$ & --              & $\pm\ 20\%$  \\
\noalign{\smallskip}\hline\noalign{\smallskip}
\end{tabular}
\end{center}
\end{table}
 
\subsection{Uncertainties in an alternative background estimate}
\label{subsec:combunc}
 
The systematic uncertainties in the alternative background estimate described in Section~\ref{subsec:alternativebkg} are evaluated as follows: the uncertainty associated with the contribution from accidental crossings is evaluated by comparing the extracted crossing factors and mass templates with those extracted from $\Lambda \rightarrow p \pi$ decays, and altering the jet selection when deriving the crossing factors. The largest difference between the resulting background estimate is taken as an uncertainty. The resulting uncertainty in the predicted accidental crossing contribution is 25\% in both of the signal regions.
 
The estimate of the hadronic interaction background relies on the assumption that the related DV mass spectrum follows an exponential shape.
In order to determine the uncertainty in the estimate via the functional fit method, a multivariate Gaussian distribution of the fit parameters is created from the covariance matrix of the fit. Sets of parameters are randomly sampled from this distribution to create alternate curves. The uncertainty is determined as the standard deviation of the envelope of the alternate curves. This is added in quadrature with the uncertainty on the shape correction originating from statistical variations in the simulated multijet events. The uncertainty amounts to a 300\% (30\%) uncertainty on the predicted hadronic interaction yield in the High-$\pt$ (Trackless) jet SR. The larger uncertainty on the estimate in the High-$\pt$ jet SR is driven by the uncertainty in the fit parameters.
 
The uncertainty on the merging rate due to statistical fluctuations is propagated to the mass template and amounts to 40\% (10\%) for the High-$\pt$ (Trackless) jet SR. Furthermore, the difference between two sets of templates with and without the $\Delta\phi$ selection and the nominal is taken as a systematic uncertainty. In the $\mDV>10$ \GeV\ region, this uncertainty is 7.4\% (20\%) for 4-trk DVs and 15\% (28\%) for $\geq 5$-trk DVs in events passing the Trackless (High-$\pt$) jet SR.
 
The uncertainties in each component are assumed to be uncorrelated and are added in quadrature to produce the final uncertainty in the total background estimate.
 
\subsection{Uncertainties on expected signal yield}
 
The uncertainty of the event yields predicted for signal scenarios is affected by several sources: uncertainties in tracking and vertex reconstruction, uncertainty in the ISR modelling in MC simulation, the jet energy scale and resolution, and uncertainty in the integrated luminosity of the data sample.
 
The uncertainty on the track reconstruction efficiency of standard tracking is determined to be approximately 1.7\%~\cite{ATL-PHYS-PUB-2015-051}. In order to estimate the uncertainty on the track reconstruction efficiency of LRT, the radial distributions of secondary vertices from $\Kshort$ decays are compared in data and Pythia multĳet simulation for each data-taking year, after normalizing the two distributions by the number of vertices at low radii, where tracking efficiencies are well understood. The mismodelling of the reconstruction efficiency in MC simulation at large track impact parameter and large vertex radius is estimated in this comparison. The observed difference is added in quadrature with the standard track reconstruction efficiency uncertainty and then applied as a conservative per-track efficiency uncertainty to the tracks associated with displaced vertices in the signal samples. The effect on final event counts with a fully selected DV is determined to be between 14\% and 17\%.
 
The effect on the signal efficiency due to variations of the amount of simulated pileup interactions is negligible. In order to estimate the size of the uncertainty due to the QCD radiation modelling, the factorization, renormalization, and merging scales were varied in \MGNLO, as well as in the parton showering tunes and radiation tunes. This effect corresponds to an uncertainty of about 20\% in the signal selection efficiency for electroweakino pair production signals with $m(\ninoone)<500$~GeV and the uncertainty decreases for higher $m(\ninoone)$. The uncertainty was found to be negligible for the gluino pair production signals. The differences between results from the nominal and varied predictions samples are used as the systematic uncertainties.
 
Uncertainties in the jet energy scale and resolution for jets originating from displaced decays were studied with the help of the simulated signal samples. They were found to impact the event counts of the EWK RPV signal model up to 10\% and were negligible for the Strong RPV signal model.
 
The impact of the systematic uncertainties affecting the signal yield in the two signal regions is summarized in \Tab{\ref{tab:sigsystsummary}}.
 
\begin{table}[!ht]
\caption{Summary of the impact of systematic uncertainties in the predicted yield for the signal scenarios considered. These uncertainties apply to both SRs used in the search.}
\label{tab:sigsystsummary}
\begin{center}
\begin{tabular}{l|p{3cm}p{3cm}}
\noalign{\smallskip}\hline\noalign{\smallskip}
Source of uncertainty              & \multicolumn{2}{c}{Relative impact signal yield [\%]}                                            \\
&                                                       &                                          \\[-1em]
& \multicolumn{1}{c}{Gluino pairs}                      & \multicolumn{1}{c}{Electroweakino pairs} \\
\noalign{\smallskip}\hline\noalign{\smallskip}
Total                              & \multicolumn{1}{c}{17 -- 20}                          & \multicolumn{1}{c}{20 -- 31}             \\
\noalign{\smallskip}\hline\noalign{\smallskip}
Tracking and vertex reconstruction & \multicolumn{2}{c}{14 -- 17}                                                                     \\
ISR modelling in MC simulation      & \multicolumn{1}{c}{$<1$}                              & \multicolumn{1}{c}{1 -- 24}              \\
Jet energy scale and resolution    & \multicolumn{1}{c}{$1$}                               & \multicolumn{1}{c}{10}                   \\
Integrated luminosity of data sample   & \multicolumn{2}{c}{1.7}                                                                          \\
\noalign{\smallskip}\hline\noalign{\smallskip}
\end{tabular}
\end{center}
\end{table}
 
\FloatBarrier
\section{Results}
 
No events are observed in the Trackless jet SR, while a single event is observed in the High-\pT\ jet SR. The observed event yields are in good agreement with the background-only expectations.
The single event observed in the High-\pT\ jet SR contains seven selected jets with \pT\ above 90.8~\GeV, and one selected DV. The DV has five tracks, $\mDV= 32.6$~\GeV, and net electric charge of $-1e$.
 
Figure~\ref{fig::2Dmaps} shows the two-dimensional distributions of \mDV\ and $\nTracks$ in events passing the High-\pT\ and Trackless jet selections in data and signal MC.
\begin{figure}[h]
\centering
\begin{subfigure}[b]{0.48\textwidth}
\includegraphics[width=\textwidth]{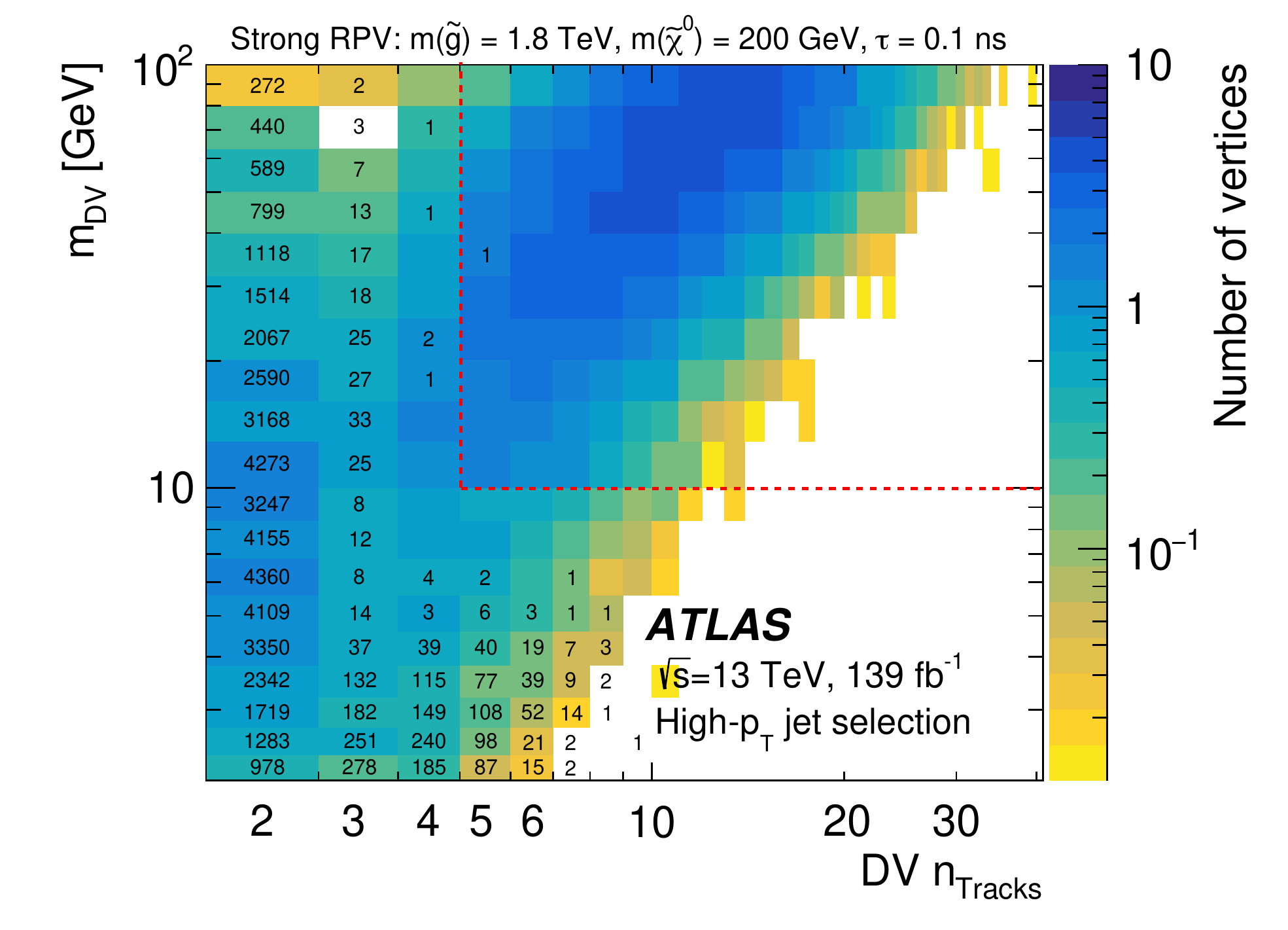}
\caption{}
\end{subfigure}
\begin{subfigure}[b]{0.48\textwidth}
\includegraphics[width=\textwidth]{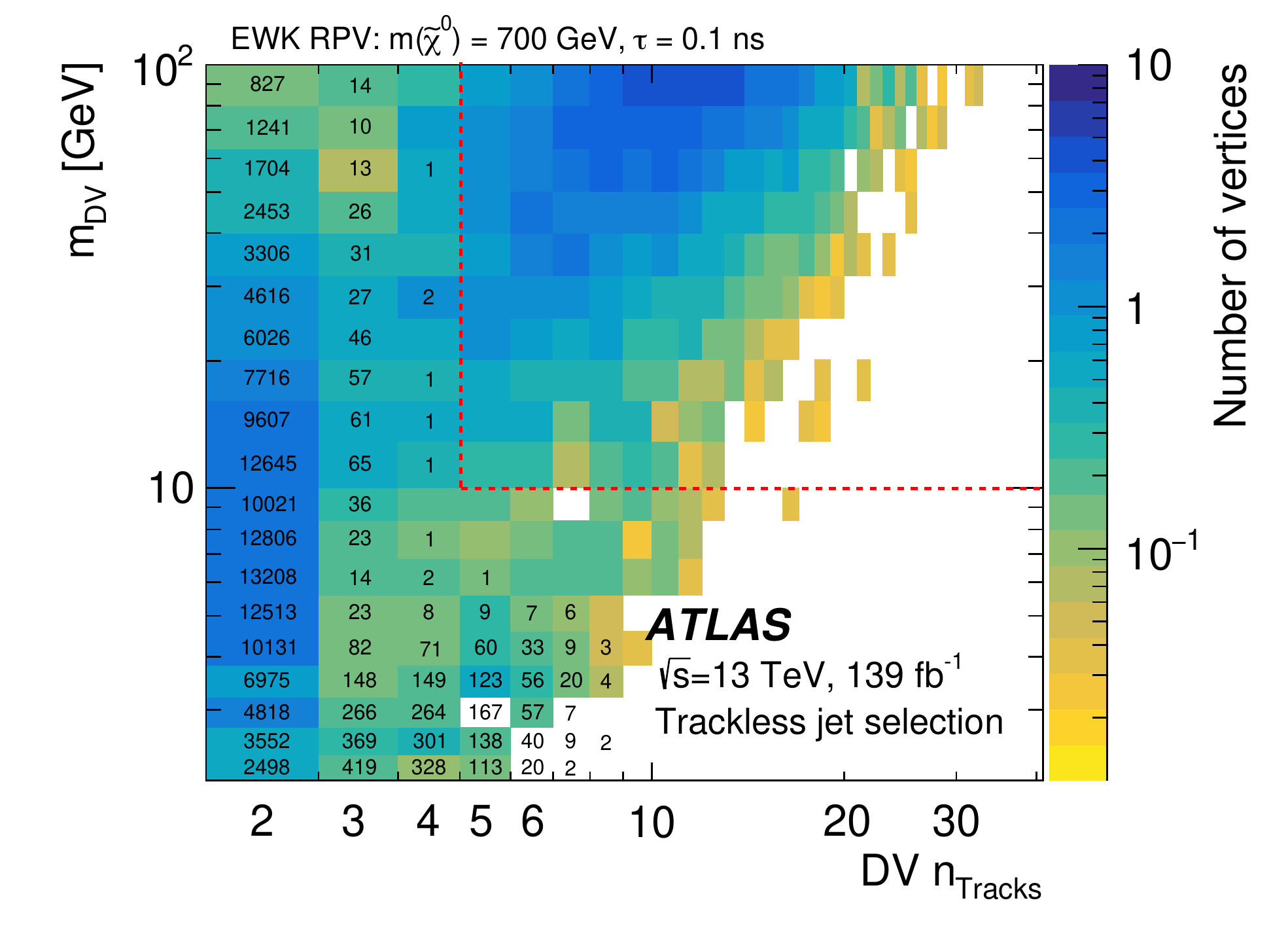}
\caption{}
\end{subfigure}
\caption{
Two-dimensional distribution of the invariant mass $\mDV$ and the track multiplicity for preselected DVs in events which satisfy the (left) High-$\pt$ and (right) Trackless jet selection requirements in data and signal MC. The shown numbers correspond to the observed vertex yields in data, while the colour-representation shows the expected signal yield of two benchmark models. The dashed line represents the boundary of the signal region requirements.}
\label{fig::2Dmaps}
\end{figure}
 
The predicted backgrounds and observed yields in the two signal regions are used simultaneously to set exclusion limits on the BSM signal models of interest. The exclusion limits are set using the $\mathrm{CL}_{s}$ prescription~\cite{Read:2002hq}.
The exclusion limit setting is implemented in \texttt{pyhf}~0.6.3~\cite{pyhf_joss, Cowan:2010js}. It is done using a two-bin fit, where the two bins correspond to the  High-$\pT$ and Trackless jet signal region, respectively. The uncertainties in the background prediction in the two SRs are considered fully uncorrelated. The uncertainties in the signal yields are taken as fully correlated.
 
The expected and observed exclusion limits with 95\% confidence level (CL) on the mass of a long-lived pure-higgsino electroweakino triplet as a function of $\tau$ are shown in Figure~\ref{fig:exclusion_C1N2N1}. The expected and observed 95\% CL exclusion limits on the mass and mean proper lifetime $\tau$ of a long-lived $\ninoone$ in gluino pair production events for a fixed $m(\gluino)=2400$ \GeV\ are shown in Figure~\ref{fig:exclusion_GG}.
For a mean proper lifetime of 0.1~ns, masses below $m(\ninoone) \simeq 1.58$~\TeV\ are excluded independently of the presence of a heavier gluino. The observed limit extends above $m(\ninoone) = 1.5$~\TeV\ for a range of mean proper lifetimes between 0.03~ns and 1~ns. Stronger limits can be set when the $\ninoone$ is produced in the decay of a heavier $\gluino$. For $\ninoone$ masses on the order of few hundred GeV and below, the sensitivity decreases because of the sizeable Lorentz boost and the limited fiducial volume.

\begin{figure}[htbp]
\centering
\includegraphics[width=0.63\textwidth]{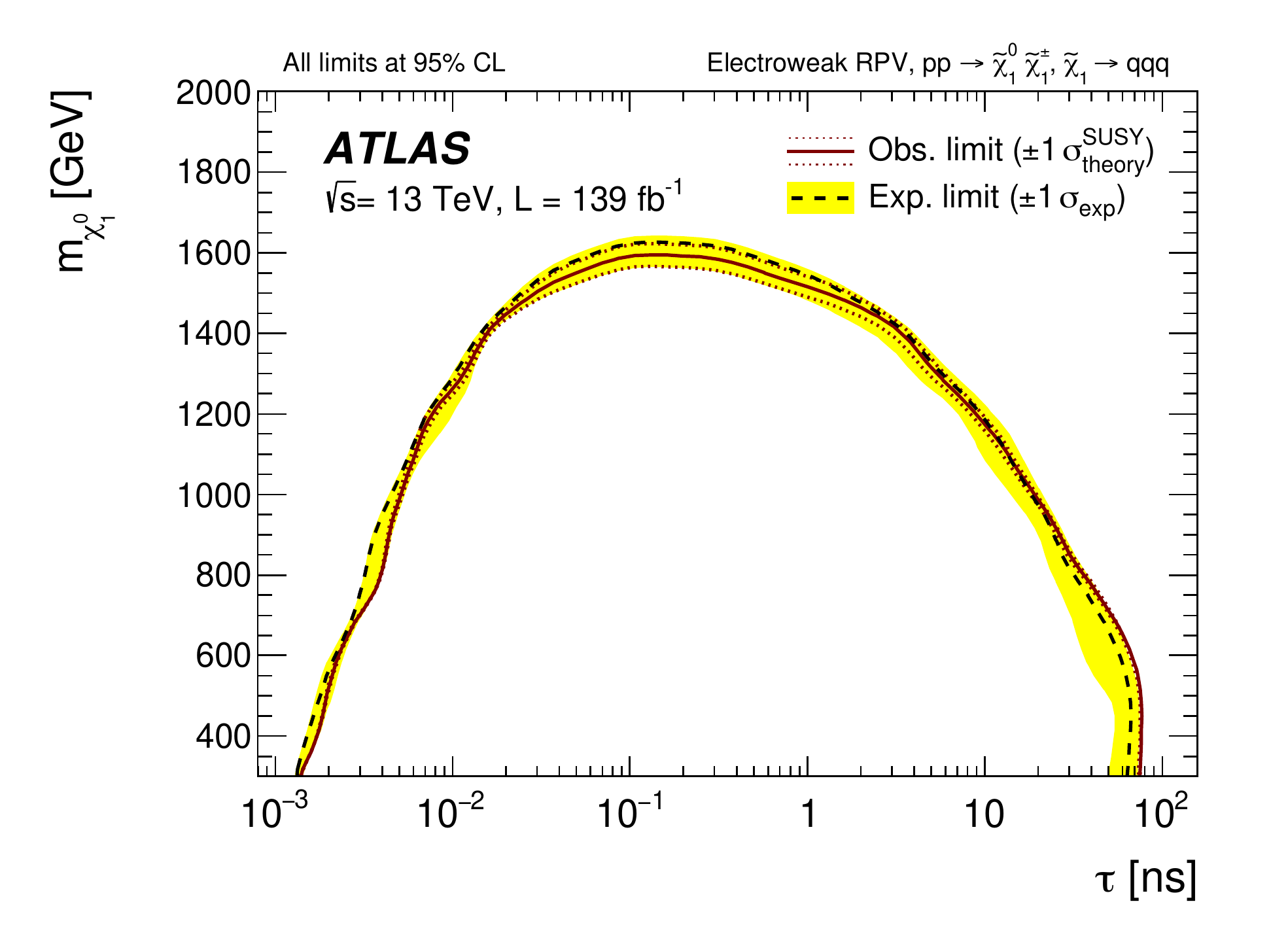}
\caption{Exclusion limits at 95\% CL on the lifetime and mass of the \ninoone in electroweakino pair production models. The dashed line and the shaded band are the expected limit and its $\pm 1 \sigma$ uncertainty, respectively. The thick solid line is the observed limit for the central value of the signal cross-section. The expected and observed limits do not include the effect of the theoretical uncertainties in the signal cross-section. The dotted lines show the effect on the observed limit of varying the signal cross-section by $\pm 1 \sigma$ of the theoretical uncertainty. The area below the solid line is the excluded parameter space.}
\label{fig:exclusion_C1N2N1}
\end{figure}
\begin{figure}[htbp]
\centering
\includegraphics[width=0.63\textwidth]{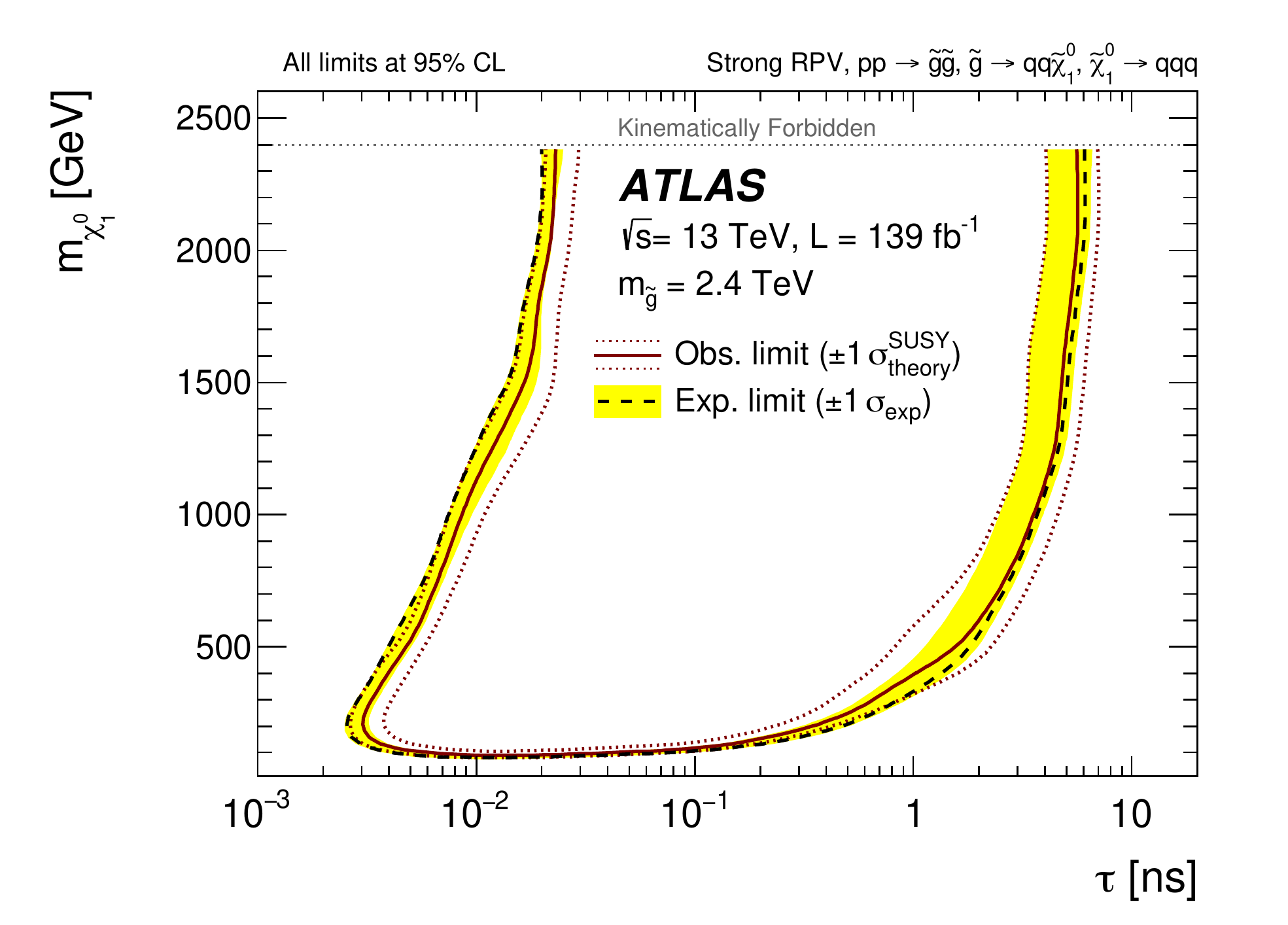}\\
\caption{Exclusion limits at 95\% CL on the lifetime and mass of the \ninoone\ in gluino pair production models for a fixed m(\gluino) = 2400~\GeV. The dashed line and the shaded band are the expected limit and its $\pm 1 \sigma$ uncertainty, respectively. The thick solid line is the observed limit for the central value of the signal cross-section. The expected and observed limits do not include the effect of the theoretical uncertainties in the signal cross-section. The dotted lines show the effect on the observed limit of varying the signal cross-section by $\pm 1 \sigma$ of the theoretical uncertainty. The area between the two solid lines is the excluded parameter space.}
\label{fig:exclusion_GG}
\end{figure}
 
In Figure~\ref{fig:xs_lims}, the 95\% CL upper limits on the production cross-section of electroweakino pairs in the EWK RPV model and gluino pairs in the Strong RPV model are shown as a function of $m(\ninoone)$. The limited variation as a function of $m(\ninoone)$ in the EWK RPV model is because the signal acceptance and efficiency are relatively invariant in $m(\ninoone)$ for the range considered here. For $\tau = 0.1$~ns, cross-section upper limits set are below 40~ab in the Strong RPV model and below 70~ab in the EWK RPV model.
 
\begin{figure}[htbp]
\centering
\begin{subfigure}[b]{0.48\textwidth}
\includegraphics[width=\textwidth]{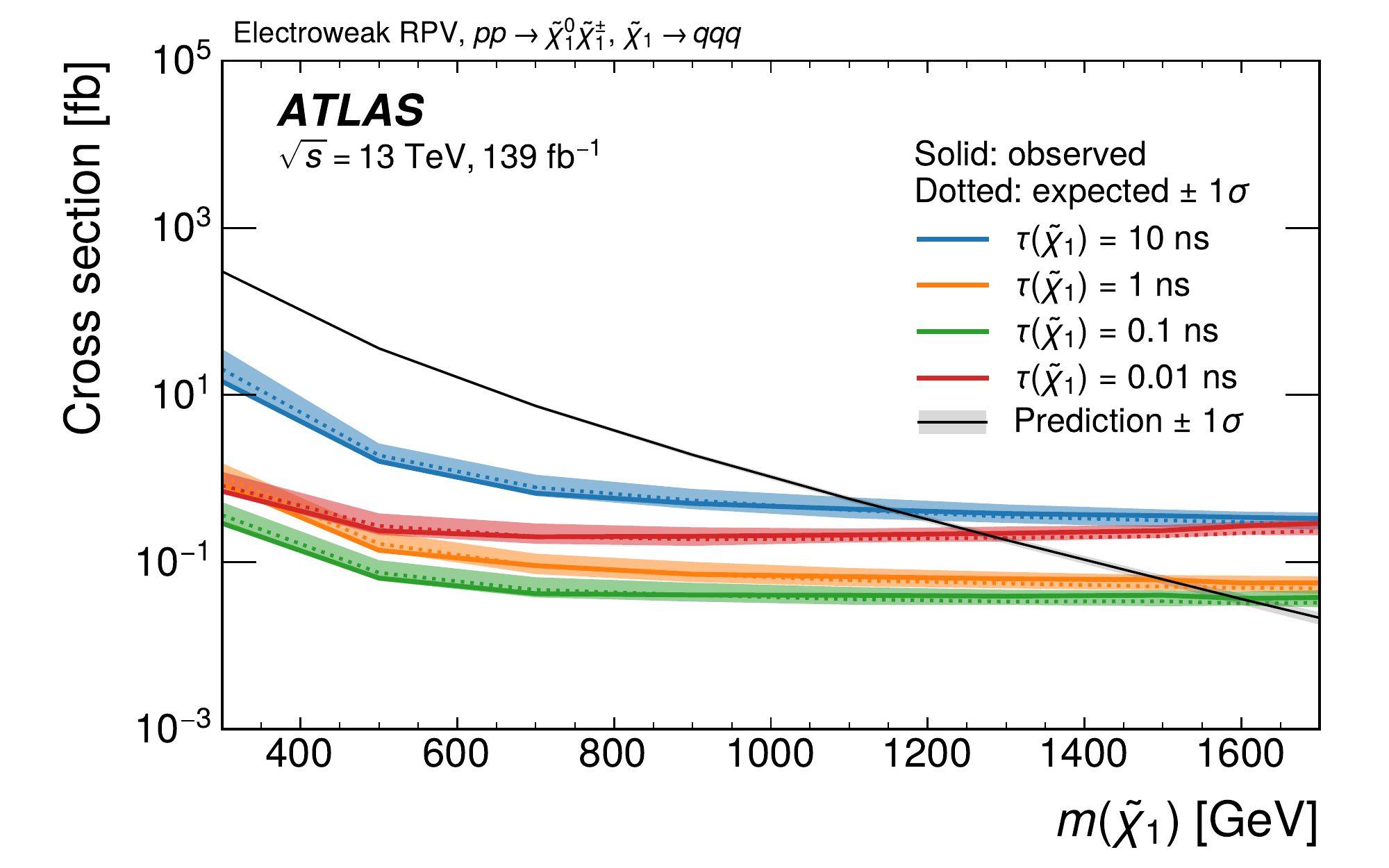}
\caption{}
\end{subfigure}
\begin{subfigure}[b]{0.48\textwidth}
\includegraphics[width=\textwidth]{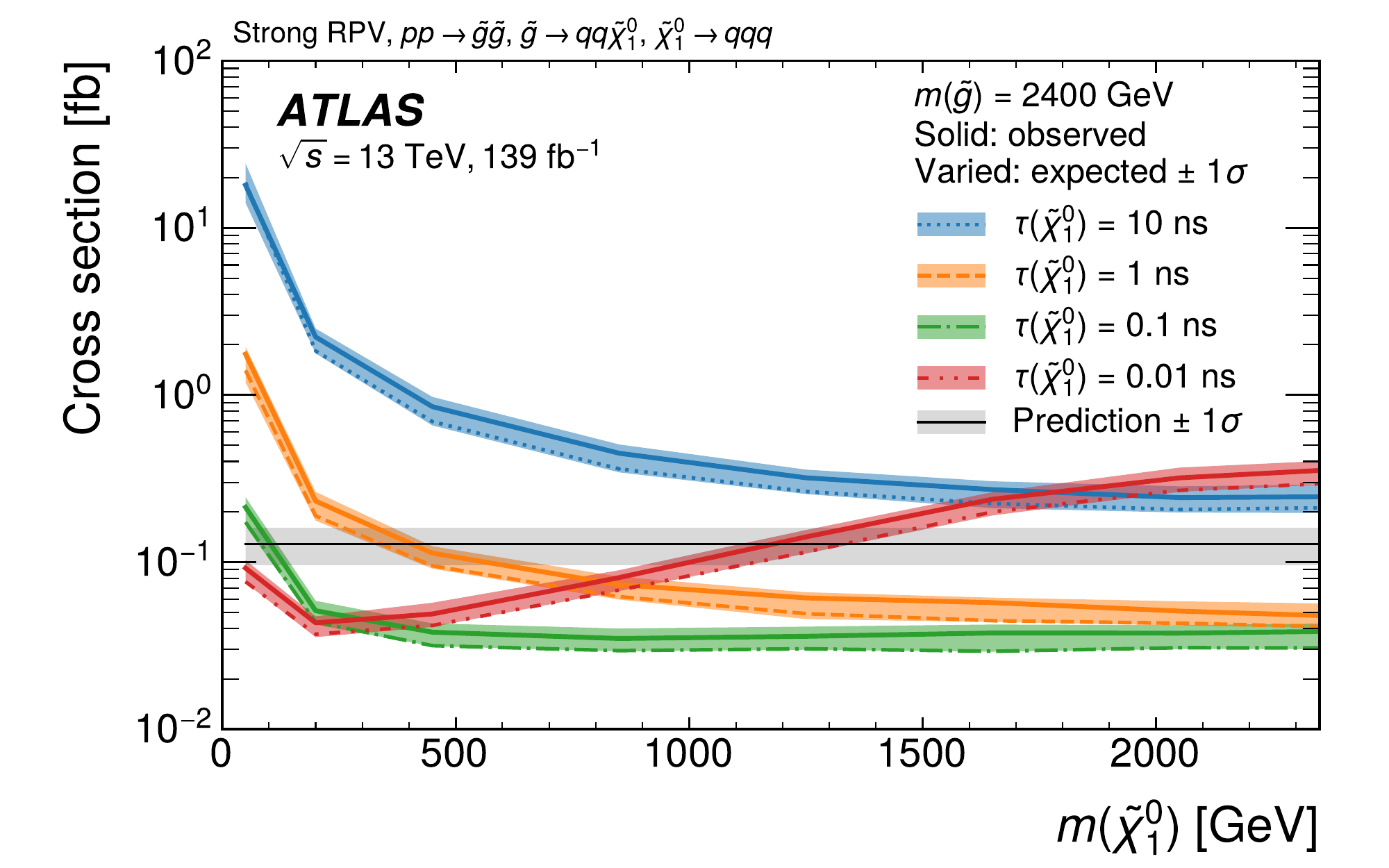}
\caption{}
\end{subfigure}
\caption{Exclusion limits at 95\% CL on the production cross section of (a) electroweakino pairs in the EWK RPV model and (b) gluino pairs in the Strong RPV model as a function of $m(\ninoone)$ are shown for several values of $\tau(\ninoone)$. The theoretical prediction of the signal production cross section and its uncertainty are shown in gray.}
\label{fig:xs_lims}
\end{figure}
 
Model-independent upper limits at 95\% CL on the number of BSM events in the signal region are also derived, assuming no significant contamination from alternate signal models in the CRs used to derive the background predictions. Normalizing these limits by the integrated luminosity of the data sample, these numbers can be interpreted as upper limits on the visible BSM cross-section, denoted by $\sigma_{\mathrm{vis}}$. This is defined as the product of signal acceptance, reconstruction efficiency, and production cross-section, and the results are given in \Tab{\ref{table.results.exclxsec.pval.upperlimit.SS}}. Parameterized signal acceptances and efficiences for the purpose of reinterpreting the results of this search are provided in a separate document~\cite{Hepdata}.
 
\begin{table}[h]
\caption[Breakdown of upper limits.]{The observed data, expected background, observed ($S_{\textrm{obs} }^{95}$) and expected ($S_{\textrm{exp} }^{95}$) limits on the number of signal events, and 95\% CL upper limits on the visible cross section $\langle\sigma_{\mathrm{vis}}\rangle_{\textrm{obs} }^{95}$.}
\centering
\setlength{\tabcolsep}{0.0pc}
\begin{tabular*}{0.7\textwidth}{@{\extracolsep{\fill}}lccccccc}
\noalign{\smallskip}\hline\noalign{\smallskip}
Signal Region & Observed & Expected &  $S_{\textrm{obs} }^{95}$  & $S_{\textrm{exp} }^{95}$ & $\langle\sigma_{\mathrm{vis}}\rangle_{\textrm{obs} }^{95}$~[fb] \\ \\[-1em]
\noalign{\smallskip}\hline\noalign{\smallskip}
High-\pT\ jet SR   &1 &$0.46^{+0.27}_{-0.30}$ & $3.8$ & ${ 3.1 }^{ +1.0 }_{ -0.1 }$ & $0.027$ \\  \\[-1em]
Trackless jet SR   &0 &$0.83^{+0.51}_{-0.53}$ & $3.0$ & ${ 3.4 }^{ +1.3 }_{ -0.3 }$ & $0.022$ \\%
\noalign{\smallskip}\hline\noalign{\smallskip}
\end{tabular*}
\label{table.results.exclxsec.pval.upperlimit.SS}
\end{table}

\FloatBarrier
\section{Conclusion}
 
A search for physics beyond the Standard Model giving rise to long-lived particle decays into hadrons is reported using the ATLAS experiment at the LHC using an accumulated data collection of 139~\ifb\ of $\sqrt{s} = 13$~\TeV\ pp collisions. Event selections are developed to efficiently reject backgrounds. The yields expected from background in the two orthogonal signal regions used in the analysis are extracted from data. The data agree with the yields expected from the background-only hypothesis, with zero and one event passing the trackless and High-\pT\ jet SR requirements, respectively.
 
The results are interpreted in supersymmetric models with electroweakinos decaying via small values of the $R$-parity-violating couplings \lampp\ into quarks, giving the electroweakinos lifetimes $\tau$  in the ps to ns range. At 95\% confidence level, $m(\ninoone)$ values up to 1.58~\TeV\ for $\tau = 0.1$~ns are excluded, independently of the presence of a heavier gluino. Upper limits on the visible cross-section for processes with the sought signature were also derived to be 0.03~fb and 0.02~fb for the High-\pT\ jet and Trackless jet SRs, respectively.
 
\section*{Acknowledgements}


We thank CERN for the very successful operation of the LHC, as well as the
support staff from our institutions without whom ATLAS could not be
operated efficiently.
 
We acknowledge the support of
ANPCyT, Argentina;
YerPhI, Armenia;
ARC, Australia;
BMWFW and FWF, Austria;
ANAS, Azerbaijan;
CNPq and FAPESP, Brazil;
NSERC, NRC and CFI, Canada;
CERN;
ANID, Chile;
CAS, MOST and NSFC, China;
Minciencias, Colombia;
MEYS CR, Czech Republic;
DNRF and DNSRC, Denmark;
IN2P3-CNRS and CEA-DRF/IRFU, France;
SRNSFG, Georgia;
BMBF, HGF and MPG, Germany;
GSRI, Greece;
RGC and Hong Kong SAR, China;
ISF and Benoziyo Center, Israel;
INFN, Italy;
MEXT and JSPS, Japan;
CNRST, Morocco;
NWO, Netherlands;
RCN, Norway;
MEiN, Poland;
FCT, Portugal;
MNE/IFA, Romania;
MESTD, Serbia;
MSSR, Slovakia;
ARRS and MIZ\v{S}, Slovenia;
DSI/NRF, South Africa;
MICINN, Spain;
SRC and Wallenberg Foundation, Sweden;
SERI, SNSF and Cantons of Bern and Geneva, Switzerland;
MOST, Taiwan;
TENMAK, T\"urkiye;
STFC, United Kingdom;
DOE and NSF, United States of America.
In addition, individual groups and members have received support from
BCKDF, CANARIE, Compute Canada and CRC, Canada;
PRIMUS 21/SCI/017 and UNCE SCI/013, Czech Republic;
COST, ERC, ERDF, Horizon 2020 and Marie Sk{\l}odowska-Curie Actions, European Union;
Investissements d'Avenir Labex, Investissements d'Avenir Idex and ANR, France;
DFG and AvH Foundation, Germany;
Herakleitos, Thales and Aristeia programmes co-financed by EU-ESF and the Greek NSRF, Greece;
BSF-NSF and MINERVA, Israel;
Norwegian Financial Mechanism 2014-2021, Norway;
NCN and NAWA, Poland;
La Caixa Banking Foundation, CERCA Programme Generalitat de Catalunya and PROMETEO and GenT Programmes Generalitat Valenciana, Spain;
G\"{o}ran Gustafssons Stiftelse, Sweden;
The Royal Society and Leverhulme Trust, United Kingdom.
 
The crucial computing support from all WLCG partners is acknowledged gratefully, in particular from CERN, the ATLAS Tier-1 facilities at TRIUMF (Canada), NDGF (Denmark, Norway, Sweden), CC-IN2P3 (France), KIT/GridKA (Germany), INFN-CNAF (Italy), NL-T1 (Netherlands), PIC (Spain), ASGC (Taiwan), RAL (UK) and BNL (USA), the Tier-2 facilities worldwide and large non-WLCG resource providers. Major contributors of computing resources are listed in Ref.~\cite{ATL-SOFT-PUB-2021-003}.


\printbibliography
 
\clearpage
 
\begin{flushleft}
\hypersetup{urlcolor=black}
{\Large The ATLAS Collaboration}

\bigskip

\AtlasOrcid[0000-0002-6665-4934]{G.~Aad}$^\textrm{\scriptsize 101}$,
\AtlasOrcid[0000-0002-5888-2734]{B.~Abbott}$^\textrm{\scriptsize 119}$,
\AtlasOrcid[0000-0002-7248-3203]{D.C.~Abbott}$^\textrm{\scriptsize 102}$,
\AtlasOrcid[0000-0002-1002-1652]{K.~Abeling}$^\textrm{\scriptsize 55}$,
\AtlasOrcid[0000-0002-8496-9294]{S.H.~Abidi}$^\textrm{\scriptsize 29}$,
\AtlasOrcid[0000-0002-9987-2292]{A.~Aboulhorma}$^\textrm{\scriptsize 35e}$,
\AtlasOrcid[0000-0001-5329-6640]{H.~Abramowicz}$^\textrm{\scriptsize 150}$,
\AtlasOrcid[0000-0002-1599-2896]{H.~Abreu}$^\textrm{\scriptsize 149}$,
\AtlasOrcid[0000-0003-0403-3697]{Y.~Abulaiti}$^\textrm{\scriptsize 116}$,
\AtlasOrcid[0000-0003-0762-7204]{A.C.~Abusleme~Hoffman}$^\textrm{\scriptsize 136a}$
% ...
\bigskip
\\

% ...
$^{29}$Physics Department, Brookhaven National Laboratory, Upton NY; United States of America.\\
% ...
$^{35}$$^{(a)}$Facult\'e des Sciences Ain Chock, R\'eseau Universitaire de Physique des Hautes Energies - Universit\'e Hassan II, Casablanca;$^{(b)}$Facult\'{e} des Sciences, Universit\'{e} Ibn-Tofail, K\'{e}nitra;$^{(c)}$Facult\'e des Sciences Semlalia, Universit\'e Cadi Ayyad, LPHEA-Marrakech;$^{(d)}$LPMR, Facult\'e des Sciences, Universit\'e Mohamed Premier, Oujda;$^{(e)}$Facult\'e des sciences, Universit\'e Mohammed V, Rabat;$^{(f)}$Institute of Applied Physics, Mohammed VI Polytechnic University, Ben Guerir; Morocco.\\
% ...
$^{55}$II. Physikalisches Institut, Georg-August-Universit\"{a}t G\"ottingen, G\"ottingen; Germany.\\
% ...
$^{101}$CPPM, Aix-Marseille Universit\'e, CNRS/IN2P3, Marseille; France.\\
$^{102}$Department of Physics, University of Massachusetts, Amherst MA; United States of America.\\
% ...
$^{116}$Department of Physics, New York University, New York NY; United States of America.\\
% ...
$^{119}$Homer L. Dodge Department of Physics and Astronomy, University of Oklahoma, Norman OK; United States of America.\\
% ...
$^{136}$$^{(a)}$Departamento de F\'isica, Pontificia Universidad Cat\'olica de Chile, Santiago;$^{(b)}$Millennium Institute for Subatomic physics at high energy frontier (SAPHIR), Santiago;$^{(c)}$Instituto de Investigaci\'on Multidisciplinario en Ciencia y Tecnolog\'ia, y Departamento de F\'isica, Universidad de La Serena;$^{(d)}$Universidad Andres Bello, Department of Physics, Santiago;$^{(e)}$Instituto de Alta Investigaci\'on, Universidad de Tarapac\'a, Arica;$^{(f)}$Departamento de F\'isica, Universidad T\'ecnica Federico Santa Mar\'ia, Valpara\'iso; Chile.\\
% ...
$^{149}$Department of Physics, Technion, Israel Institute of Technology, Haifa; Israel.\\
$^{150}$Raymond and Beverly Sackler School of Physics and Astronomy, Tel Aviv University, Tel Aviv; Israel.\\
% ...
$^{a}$ Also Affiliated with an institute covered by a cooperation agreement with CERN.\\
$^{b}$ Also at An-Najah National University, Nablus; Palestine.\\
$^{c}$ Also at Borough of Manhattan Community College, City University of New York, New York NY; United States of America.\\
$^{d}$ Also at Bruno Kessler Foundation, Trento; Italy.\\
$^{e}$ Also at Center for High Energy Physics, Peking University; China.\\
$^{f}$ Also at Center for Interdisciplinary Research and Innovation (CIRI-AUTH), Thessaloniki ; Greece.\\
% ...

\end{flushleft}


\FloatBarrier

\FloatBarrier

\end{document}